\documentclass[journal]{IEEEtran}
\usepackage{amsmath}
\usepackage{amssymb}
\usepackage{amsfonts}
\usepackage{epsfig}
\usepackage{graphicx}
\usepackage{color}
\usepackage{cite}

\title{Large Families of Optimal Two-Dimensional Optical Orthogonal Codes}
\author{Reza Omrani, Gagan Garg, P. Vijay Kumar, Petros Elia, and Pankaj Bhambhani
\thanks{The results of this paper have been presented in part at ISIT 2005, 2006 and SETA 2004, 2006.}
\thanks{Reza Omrani is with Telegent Systems, 470 Porrero Avenue,
Sunnyvale, CA 94085, USA. {\tt omrani@usc.edu}}
\thanks{Gagan Garg is with the LNM Institute of Information Technology, Jaipur 302031, India. Part of this work was carried out while he was at Indian Institute of Science. {\tt gagan.garg@gmail.com}}
\thanks{P. Vijay Kumar is currently with the Department of Electrical Communication Engineering, Indian
Institute of Science, Bangalore 560012, India. Part of this work was carried out while he was
at University of Southern California.  {\tt vijayk@usc.edu}}
\thanks{Petros Elia is with the Department of Mobile Communications, EURECOM, BP 193, F-06904,
Sophia Antipolis cedex, France. {\tt petros.elia@eurecom.fr}}
\thanks{Pankaj Bhambhani is with Cisco Systems India Pvt. Ltd.,
Cessna Business Park, Sarjapur, Bangalore 560087,
India. {\tt pankaj.ani2000@gmail.com} }
\thanks{This research was
supported in part by DARPA OCDMA Program Grant No. N66001-02-1-8939.}}

\newcommand{\beq}{\begin{equation}}
\newcommand{\eeq}{\end{equation}}
\newcommand{\beqn}{\[}
\newcommand{\eeqn}{\]}
\newcommand{\bea}{\begin{eqnarray}}
\newcommand{\eea}{\end{eqnarray}}
\newcommand{\bean}{\begin{eqnarray*}}
\newcommand{\eean}{\end{eqnarray*}}
\newcommand{\re}{\mbox{$\mathfrak{Re}$}}
\newcommand{\bit}{\begin{itemize}}
\newcommand{\eit}{\end{itemize}}

\newcommand{\ben}{\begin{enumerate}}
\newcommand{\een}{\end{enumerate}}

\newtheorem{theorem}{Theorem}

\newtheorem{note}{Remark}

\newtheorem{prop}[theorem]{Proposition}
\newtheorem{cor}[theorem]{Corollary}

\newtheorem{example}{Example}
\newtheorem{defn}{Definition}

\newcommand{\pgfq}{\mathbb{P}^1(\mathbb{F}_{q})}
\newcommand{\gfq}{\mathbb{F}_{q}}

\begin{document}

 \maketitle

\begin{abstract}

Nine new 2-D OOCs are presented here, all sharing the common feature
of a code size that is much larger in relation to the number of time
slots than those of constructions appearing previously in the
literature. Each of these constructions is either optimal or
asymptotically optimal with respect to either the original Johnson
bound or else a non-binary version of the Johnson bound introduced
in this paper.

The first 5 codes are constructed using polynomials over finite
fields - the first construction is optimal while the remaining 4 are
asymptotically optimal.  The next two codes are constructed using
rational functions in place of polynomials and these are
asymptotically optimal.  The last two codes, also asymptotically
optimal, are constructed by composing two of the above codes with a
constant weight binary code.

Also presented, is a three-dimensional OOC that exploits the
polarization dimension.

Finally, phase-encoded optical CDMA is considered and construction
of two efficient codes are provided.
    \end{abstract}

%
%

\section{Introduction}

There has been an upsurge of interest in applying code division
multiple access (CDMA) techniques to optical networks - optical CDMA
(OCDMA) \cite{Sal}. This is partly due to the increase in security
afforded by OCDMA (as measured, for instance, by the increased
effort needed to intercept an OCDMA signal) and partly due to the
flexibility and simplicity of network control afforded by OCDMA.

 There are two main approaches to data modulation and spreading in optical CDMA (OCDMA). The first
 approach, known as direct-sequence encoding \cite{Sal}, makes use
 of on-off-keying (OOK) data modulation and unipolar spreading sequences with good correlation
 properties. Traditionally, as in the case of wireless communication, the
spreading has been carried out in the time domain and we will refer
to this class of OOCs as one-dimensional OOCs (1-D OOCs)
\cite{ChuSalWei,ChuKum,NguGyoMas,MorZhaKumZin,BitEtz,YanFuj,Bur1,Yin,FujMia,GeYin,FujMiaYin,
TanYin,Bur2,ChuGol,DinXin,ChaFujMia,ChaMia,ChaJi,AbeBur,ChaYin,ChuCol1,ChuCol2,MiyMizShi,moreno2007generalized}.
A drawback of 1-D OOCs is the requirement of a large chip rate. By
employing two-dimensional optical orthogonal codes (2-D OOCs) that
spread in both time and wavelength domain, it turns out that the
large-chip-rate requirement can be substantially reduced. There is a
considerable literature on 2-D OOC constructions. However, in this
paper, we focus our attention only on optimal and asymptotically
optimal constructions. A quick overview of optimal (or
asymptotically optimal) constructions of 2-D codes in the literature
is presented in Table \ref{tab:lit}. The code constructions
presented here came about as a result of a DARPA-funded
project~\cite{DARPA}, a subgoal of which was coming up with
constructions that would give the experimentalist the maximum
possible flexibility in choosing the parameters $\Lambda$ and $T$.
In this context, Fig.~\ref{fig:results} provides a visual depiction
of the parameter sets for small $\Lambda, T$ in the range, $ 2 \leq
\Lambda \leq 17$, $2 \leq T \leq 33$ for which constructions are now
available as a result of the $9$ constructions presented here.

The second OCDMA approach uses spectral encoding. In this method,
 spreading is achieved by encoding of amplitude or phase of the data
  spectrum \cite{WeiHerSal,SalWeiHer}.

This paper is organized as follows: Section \ref{sec:background}
provides background material along with an overview of the results
of this paper. In Section \ref{sec:new_bounds}, we propose two new
bounds on the size of 2-D OOCs - we use these bounds to prove
optimality of some of the constructions presented in the current
paper as well as of two constructions previously known in the
literature, but which were not known to be optimal.  In the next
section, we propose five families of 2-D OOCs constructed using
polynomials over finite fields. In Section \ref{sec:rational}, we
present two families of asymptotically optimal codes constructed
using rational functions over finite fields. We show in Section
\ref{sec:concatenation} how one can generate two asymptotically
optimal families by composing two of the previous constructions with
a constant weight binary code. In Section \ref{sec:3D}, we present a
three-dimensional OOC using polarization as the third dimension. In
Sections \ref{sec:phase} and \ref{sec:phase_construction}, we use
generalized bent functions to construct two families of efficient
asynchronous phase-encoding sequences for Optical CDMA. The last
section concludes the paper. New results are presented as
Propositions, known results appear as Theorems. Most of the proofs
have been moved to the Appendices to ensure smooth reading of the
paper.

\begin{table*}[t]
\caption{Optimal and asymptotically optimal $2$-D optical orthogonal
codes in the literature}\label{tab:lit}

        \begin{center}
            \begin{tabular}{|l|l|l|l|c|}
                \hline \hline
                & & & &\\
                Construction &Parameters    &Code Size  & Constraint & Optimality\\
                Name & & & Satisfied &\\
                \hline
                & & & &\\
                Lee, Seo \cite{lee2002ncm}
                &$(\Lambda \times T,3,1)$
                &$ \left.
                                     \begin{array}{cc}
                                     6 \Lambda s t + \Lambda s + \Lambda t, \mbox{ where } s \mbox{ and } t \\
                                     \mbox{ are the sizes of the optimal OOCs} \\
                                     (\Lambda, 3, 1) \mbox{ and } (T, 3, 1) \mbox{ respectively}.\\
                                     \mbox{(O) when } \Lambda, T \equiv 1 \pmod{6};\\
                                     \mbox{(AO) otherwise}
                                     \end{array}
                                     \right.
                                   $ & None & (O)\\
                \hline
                & & & &\\
                Shurong & $(\Lambda\times T,\omega,1)$, &$\frac{\Lambda(\Lambda T - 1)}{\omega(\omega-1)}$ & None & (O)\\
                et. al. \cite{shurong2006nfd} & $\Lambda = p^k$ & & & \\
                \hline
                & & & & \\
                Kwong,  &$(\Lambda \times T, \omega, 1)$, &$\frac{\Lambda^2}{p_k}  + \frac{\Lambda^2}{p_k p_{k-1}} + \frac{\Lambda^2}{p_k p_{k-1} p_{k-2}} + \ldots + \Lambda   $ & AM-OPPTS & (AO)\\
                Yang \cite{KwoYan} &$\Lambda = p_1 p_2 \cdots p_k$, $T = p_1$, & & & \\
                &$ p_k \ge p_{k-1} \ge \ldots \ge p_1 \ge \omega $&  & & \\
                \hline
                &  & & & \\
                Yang,  &$(\Lambda \times T, \omega, 1),$ &$\frac{\Lambda \left( \Lambda^2 - 1 \right)}{\omega(\omega-1)}$ & None & (O)\\
                Kwong \cite{YanKwo}& $ \Lambda = T = \omega t (\omega - 1) + 1, $& & & \\
                &$\Lambda$ is prime, $t$ is some integer & & & \\
                \hline
                &  & & & \\
                Yang, &$(\Lambda \times T, \omega, 1),$ &$T$ & OPPW & (O)$^*$\\
                Kwong  \cite{YanKwo} & $  \omega = \Lambda$, $T = p_1 p_2 \cdots p_k, $ & & & \\
                &$ p_k \ge p_{k-1} \ge \ldots \ge p_1 \ge \Lambda $ &  & & \\
                \hline
                &  & & &\\
                Yang,  &$(\Lambda \times T, \omega, 1), \omega = \Lambda - 1, $ &$\frac{\Lambda T}{\Lambda - 1}$ & AM-OPPW & (AO)\\
                Kwong  \cite{YanKwo} & $ \Lambda = p_1, T = (p_1 - 1) p_2 \cdots p_k, $ & & & \\
                &$ p_k \ge p_{k-1} \ge \ldots \ge p_1$ & & & \\
                \hline
                & & & &\\
                Kwong &$(\Lambda \times T, \omega, 1)$, &$\Lambda^2 \cdot \Phi_{OOC} $, & AM-OPPTS  & (AO)\\
                et. al. \cite{KwoYanBabBrePru} &$\Lambda = p_1 p_2 \cdots p_k$, & where $\Phi_{OOC}$ is the cardinality & & \\
                &$ p_k \ge p_{k-1} \ge \ldots \ge p_1 \ge \omega $&  of the optimal $(T, \omega, 1)$ OOC & & \\
                \hline
                &  & & & \\
                Shivaleela &$(\Lambda \times T, \omega, 1),$ &$T $ & OPPW & (O)$^*$\\
                et. al. \cite{ShiSivSel}  & $ \Lambda = T = \omega$, $T$ is prime & & & \\
                \hline \hline
            \end{tabular}
            \end{center}

\begin{flushleft}
\begin{tabular}{l}
\hspace{1.5in} Here, \\
 \hspace{1.75in}  $\bullet$ (O) denotes Optimal, \\
 \hspace{1.75in}  $\bullet$ (AO) denotes Asymptotically Optimal, \\
\hspace{1.75in}  $\bullet$ $p$ or $p_i$ denotes a prime.\\
 \hspace{1in} $^*$ These constructions are shown to be optimal using the bounds proposed in this paper.
\end{tabular}
\end{flushleft}

    \end{table*}

\begin{table*}[t]
\caption{New $2$-D optical orthogonal codes proposed in this
paper}\label{2DOOC_table}

        \begin{center}
            \begin{tabular}{|c|l|l|l|c|}
                \hline \hline
                & & & &\\
                Construction          &Parameters    &Code Size & Constraint & Optimality\\
                Name & & & Satisfied &\\
                \hline
                & & & &\\
                P1 &$(\Lambda \times T, \omega, \kappa),$ &$T^{\kappa}$ & OPPW & (O)\\
                & $ \kappa < \omega = \Lambda \le T$, & & & \\
                & $ T $ is prime & & &\\
                \hline
                &  & & &\\
                P2 &$(\Lambda \times T, \omega, \kappa),$ &$\frac{1}{T} \sum_{d | (\Lambda - 1)} \left( \Lambda^{\left\lceil \frac{\kappa + 1}{d} \right\rceil} - 1  \right) \mu (d)$ & OPPTS & (AO)\\
                & $\Lambda = p$, $ \omega = T$, & & & \\
                &  $\kappa < \omega$,  $T \, | \,  p - 1$ &   & &\\
                \hline
                &  & & & \\
                P3 &$(\Lambda \times T, \omega, \kappa),$ &$\frac{(T + 1)^{\kappa + 1}-1}{T}$ & AM-OPPW & (AO)\\
                & $1 \le \Lambda \le p^m$, \ $ T = p^m -1$,& & & \\
                & $ \omega = \Lambda - \kappa$, $\kappa < \omega$ & & & \\
                \hline
                &  & & & \\
                P4 &$(\Lambda \times T, \omega, \kappa),$ &$\frac{1}{T} \sum_{d | \Lambda} \left( (\Lambda + 1 )^{\left\lceil \frac{\kappa + 1}{d} \right\rceil} - 1  \right) \mu(d)$ & AM-OPPTS & (AO)\\
                & $ \omega = T - \kappa$, $T \, | \,  p^m-1$, & & & \\
                & $\Lambda = p^m - 1 $, $\kappa < \omega$ & & & \\
                \hline
                &  & & & \\
                P5 &$(\Lambda \times T, \omega, \kappa),$ &$\frac{1}{T} \sum_{d | (\Lambda - 1)} \left( \Lambda^{\left\lceil \frac{\kappa + 1}{d} \right\rceil} - 1  \right) \mu(d)$ & OPPTS & (AO)\\
                & $ \omega = T$, $\Lambda = p^m $,& & & \\
                & $\kappa < \omega $,  $T \, | \, p^m - 1$ & & &\\
                \hline
                &  & & & \\
                R1 &$(\Lambda \times T, \omega, \kappa),$ &$\frac{c \left( \frac{\kappa}{2} \right) }{T} + 1 $ & OPPW & (AO)\\
                & $ \omega = \Lambda $, $\Lambda \le T - 1$, & & & \\
                & $T = p^m + 1 $, $\kappa < \omega$ is even & & &\\
                \hline
                &  & & & \\
                R2 &$(\Lambda \times T, \omega, \kappa),$ & $\frac{1}{T(q-1)} \sum_{h(x)}  u(d-deg(h(x)),T,1)  \hat{\mu}(h(x))$ & OPPTS & (AO)\\
                &$T | p^m-1 $, $\Lambda = p^m + 1$,& where the sum is over monic $h(x) \in {\mathbb F}_q[x]$ & &\\
                & $ \omega = T$,  $\kappa<\omega$ is even & of deg $\leq d$ & &\\
                \hline
                &  & & &\\
                CP1 &$(\Lambda \times T, \omega, \kappa)$, & $T^{\kappa} \left\lfloor \frac{\Lambda}{\omega}  \left\lfloor \frac{\Lambda - 1}{\omega - 1 }  \left\lfloor \frac{\Lambda - 2}{\omega - 2}  \ldots  \left\lfloor \frac{\Lambda -\kappa}{\omega - \kappa} \right\rfloor  \right\rfloor  \right\rfloor  \right\rfloor $ & AM-OPPW & (AO)\\
                &$\kappa<\omega \le \Lambda$, $T$ is prime&  & & \\
                \hline
                &  & & & \\
                CR1 &$(\Lambda \times T, \omega, \kappa)$, & $\left(\frac{c \left( \frac{\kappa}{2} \right) }{T} + 1 \right) \left\lfloor \frac{\Lambda}{\omega}  \left\lfloor \frac{\Lambda - 1}{\omega - 1 }  \left\lfloor \frac{\Lambda - 2}{\omega - 2}  \ldots  \left\lfloor \frac{\Lambda -\kappa}{\omega - \kappa} \right\rfloor  \right\rfloor  \right\rfloor  \right\rfloor $ & AM-OPPW & (AO)\\
                &$\omega \le \Lambda$, $\kappa<\omega$ is even,  & & & \\
                & $T = p^m + 1 $&  & & \\
                \hline \hline
\end{tabular}
            \end{center}
\begin{flushleft}
\begin{tabular}{l}
\hspace{1.5in} Here, \\
 \hspace{1.75in}  $\bullet$ (O) denotes Optimal and (AO) denotes Asymptotically Optimal, \\
 \hspace{1.75in}  $\bullet$ $p$ denotes a prime, $q=p^m$ \\
 \hspace{1.75in} $\bullet$ $\mu(\cdot)$ is the Mobius function, \\
 \hspace{1.75in} $\bullet$ $\hat{\mu}(\cdot)$ and $u(\cdot)$ are defined in equations  \eqref{eq:mu-hat} and \eqref{eq:r3_u_all} respectively, and \\
 \hspace{1.75in} $\bullet$ $c(t)=$ $ \left\{
                                     \begin{array}{cc}
                                     q^{2t+1}-q, &t=1,2,3,4,5,6 \\
                                     \geq  q^{2t+1}-\frac{q^{2t-6}}{7}, &t \geq 7.
                                     \end{array}
                                     \right.$
\end{tabular}
\end{flushleft}

    \end{table*}


\section{Background and Results}\label{sec:background}

The focus of the entire paper (except for Sections \ref{sec:phase}
and \ref{sec:phase_construction}) is on direct-sequence encoding.
Phase-encoding is restricted to Sections \ref{sec:phase} and
\ref{sec:phase_construction} only.

The advent of Wavelength-Division-Multiplexing (WDM) and dense-WDM
(D-WDM) technology has made it possible to spread in both wavelength
and time \cite{YanKwo}. The corresponding codes are variously called
wavelength-time hopping codes and multiple-wavelength codes. Here,
we will simply refer to these codes as two-dimensional OOCs (2-D
OOCs).

A 2-D $(\Lambda\times T,\omega,\kappa)$ OOC  $\cal{C}$ is a family
of $\{0,1\}$ $\Lambda \times T$ arrays of constant weight $\omega$.
Every pair $\{A,B\}$ of arrays in $\cal{C}$ is required to satisfy:
        \begin{equation}\label{2-D OOc}
        \sum_{\lambda=1}^{\Lambda}\sum_{t=0}^{T-1}A(\lambda,t)B(\lambda,(t\oplus_T\tau))
        \leq \kappa,
        \end{equation}
where either $A \neq B$ or $\tau \neq 0$.  We will refer to $\kappa$
as the maximum collision parameter (MCP) when in addition to
\eqref{2-D OOc} holding for all $\tau$, we have that equality holds
in \eqref{2-D OOc} for some pair $A, B$ and for some $\tau$. Note
that the asynchronism is present only along the time axis.

%
%
%
%

 \begin{figure}[h]
\begin{center}
        \includegraphics[width=3in]{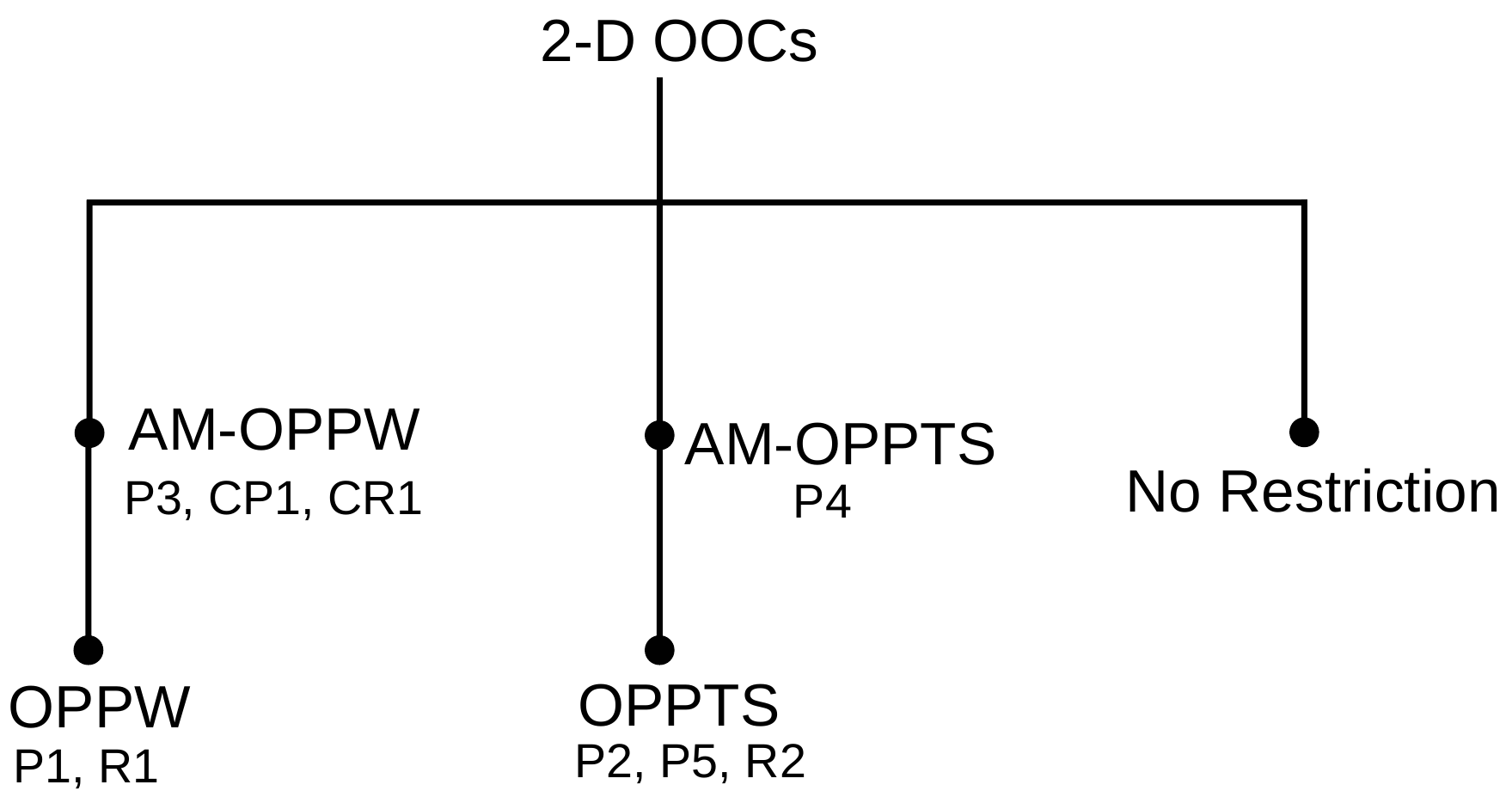}

\vspace*{0.1in}

        \caption{Various types of 2-D OOCs. The symbols P3, CP1 etc. are reference to specific
        constructions appearing in Table \ref{2DOOC_table}. }\label{fig:tree}
        \end{center}
\end{figure}

Practical considerations often place restrictions on the placement
of pulses within an array.  With this in mind, we introduce the
following terminology (see Fig. \ref{fig:tree}): \begin{itemize}
\item arrays with one-pulse per wavelength (OPPW): \ \ each row of
every $(\Lambda \times T)$ code array in ${\cal C}$ is required to
have Hamming weight $=1$. \item arrays with at most one-pulse per
wavelength (AM-OPPW): \ \ each row of any $(\Lambda \times T)$ code
in ${\cal C}$ is required to have Hamming weight $\leq 1$.
\item arrays with one-pulse per time slot (OPPTS): \ \ each
column of every $(\Lambda \times T)$ code array in ${\cal C}$ is
required to have Hamming weight $=1$. \item arrays with at most
one-pulse per time slot (AM-OPPTS): \ \ each column of any
$(\Lambda \times T)$ array in ${\cal C}$ is required to have
Hamming weight $\leq 1$ .

\end{itemize}

The constructions mentioned in Fig. \ref{fig:tree} are proposed in this paper and have been summarized in Table \ref{2DOOC_table}.

\begin{note} \label{note:autocorr}
Note that for codes that are AM-OPPW or OPPW, the autocorrelation
for non-zero values of the time shift is zero. This is obvious since
there is (at most) one $1$ in each row; hence, the time-shifted code
matrix cannot have any overlap with the original code matrix. We
shall use this fact later while proving the correlation properties
of our constructions.
\end{note}

%
%
%

\subsection{Johnson Bound} \label{sub:johnson}

For a given set of values of $\Lambda, T, \omega, \kappa$, let
$\Phi(\Lambda \times T,\omega, \kappa)$ denote the largest possible
cardinality of a $(\Lambda \times T, \omega, \kappa)$ 2-D OOC. The
following adaptation of the Johnson's bound for constant weight
codes to 2-D OOCs was first noted by Yang and Kwong in
\cite{YanKwo}:

 \begin{theorem}  \label{thm:2-D_bound_A} [Johnson Bound] \beq \label{2-D bound_A}
\Phi(\Lambda \times T,\omega,\kappa)  \leq
\left\lfloor\frac{\Lambda}{\omega}\left\lfloor\frac{\Lambda
T-1}{\omega-1}\cdots \left\lfloor\frac{\Lambda
T-\kappa}{\omega-\kappa}\right\rfloor\right\rfloor\right\rfloor .
 \eeq
 \end{theorem}

A construction meeting this bound with equality is called an optimal
construction. A construction that meets this bound asymptotically
(i.e., when $\Lambda$ or $T$ (or both) tend to infinity) is called
an asymptotically optimal construction.

\subsection{Literature Review} \label{sub:lit}
In this subsection, we focus primarily on prior constructions of
optimal or asymptotically optimal constructions of $2$-D OOCs in the
literature. A summary of these appears in Table \ref{tab:lit}. Note
that all these constructions are optimal (or asymptotically optimal)
only for MCP $=1$. However, the constructions that we propose in
this paper are optimal (or asymptotically optimal) for all values of
the MCP, i.e., MCP $\ge 1$ thereby leading to larger size (this is
explained in detail in the next subsection).

The construction by Lee and Seo \cite{lee2002ncm} spreads in the
wavelength and the time domain by using two different 1-D OOCs.
Shurong et. al \cite{shurong2006nfd} construct a 2-D OOC by
employing a frequency hopping code to spread in the wavelength
domain and a 1-D OOC to spread along the time axis. The construction
by Kwong and Yang \cite{KwoYan} interchanges the time and wavelength
components of a frequency-hopping code and then applies specific
cyclic shifts to control the value of the MCP. The first
construction by Yang and Kwong \cite{YanKwo} uses a 1-D OOC to
achieve spreading in the wavelength and time domains. The remaining
two constructions in \cite{YanKwo} modify frequency-hopping codes to
construct 2-D OOCs. The construction by Kwong et. al
\cite{KwoYanBabBrePru} spreads in the wavelength domain using a
frequency-hopping code and in the time domain using a 1-D OOC. The
OPPW construction by Shivaleela et. al \cite{ShiSivSel} places a $1$
in the first time slot of the first wavelength. By cyclically
shifting the position of the $1$ in the subsequent wavelengths by
$k$, the entire 2-D code is generated. The $k$ for different
codewords varies from $0$ to $T-1$, where $T$ is the number of time
slots.

Additional papers in the literature dealing with the design of 2-D
OOCs include
\cite{ShiSelSri,1549270,saghari2005analytical,ParMenGar,kit,TanAnd,TanAndTurBud,YanKwo1,
FatRusLar,MenGagFenHerMor,YimCheBaj,MenGagHerBenLen,liang2008nfv,
yu1999dnf, kwong2005nfw, gu2005ctd,
chang2006wtc,morelle2007doc,lin2007cap,YanKwoCha03,LemGre,SimOmuSchLev,Ein,Kum,Sar,MorMar}.
However, since the focus of the current paper is on optimal or
asymptotically optimal constructions, we do not discuss these
further here. A paper relating to 3-D code construction is
\cite{kim2000nfs}.


\subsection{Overview of Results} \label{sub:perspective}

We propose a version of non-binary Johnson bound and derive two
other bounds from it - these bounds provide upper bounds on the size
of 2-D OOCs for the case of AM-OPPW 2-D OOCs and OPPW 2-D OOCs. A
special instance is shown to lead to the Singleton bound.

We then propose 9 new families of 2-D OOCs of large size. All the
codes proposed in this paper are optimal (or asymptotically optimal)
with respect to the original Johnson bound \cite{YanKwo} or the new
bounds proposed in this paper. We obtain codes with large size by
constructing optimal families for large values of the MCP. Consider
MCP $= \kappa = 1$, for example. The 2-D Johnson bound gives \bean
\Phi(\Lambda \times T,\omega,\kappa) & \leq  &
\left\lfloor\frac{\Lambda}{\omega}\left\lfloor\frac{\Lambda
T-1}{\omega-1}\right\rfloor\right\rfloor \\
& \approx & \frac{\Lambda}{\omega}\left(\frac{\Lambda
T-1}{\omega-1}\right) \\
& \approx & \frac{\Lambda^2 T}{\omega^2} \\
& = & \frac{1}{T} \left( \frac{\Lambda T}{\omega} \right)^2.
 \eean

Similarly, for large values of $\kappa$, we get

\beqn \Phi(\Lambda \times T,\omega,\kappa) \lesssim \frac{1}{T}
\left( \frac{\Lambda T}{\omega} \right)^{\kappa + 1}. \eeqn

This shows that, for fixed values of $\Lambda, T$ and $\omega$, the
upper bound on the maximum number of codewords increases
exponentially in the MCP. Note that all the constructions in Table
\ref{tab:lit} have MCP $= 1$ thereby restricting the size to $
\frac{\Lambda^2 T}{\omega^2}$. Hence, by constructing optimal (or
asymptotically optimal) constructions for larger values of the MCP,
we are proposing 2-D OOCs with size larger than has been previously
constructed.

\begin{figure*}[ht!]
\begin{center}
        \includegraphics[angle=90, width=6.4in]{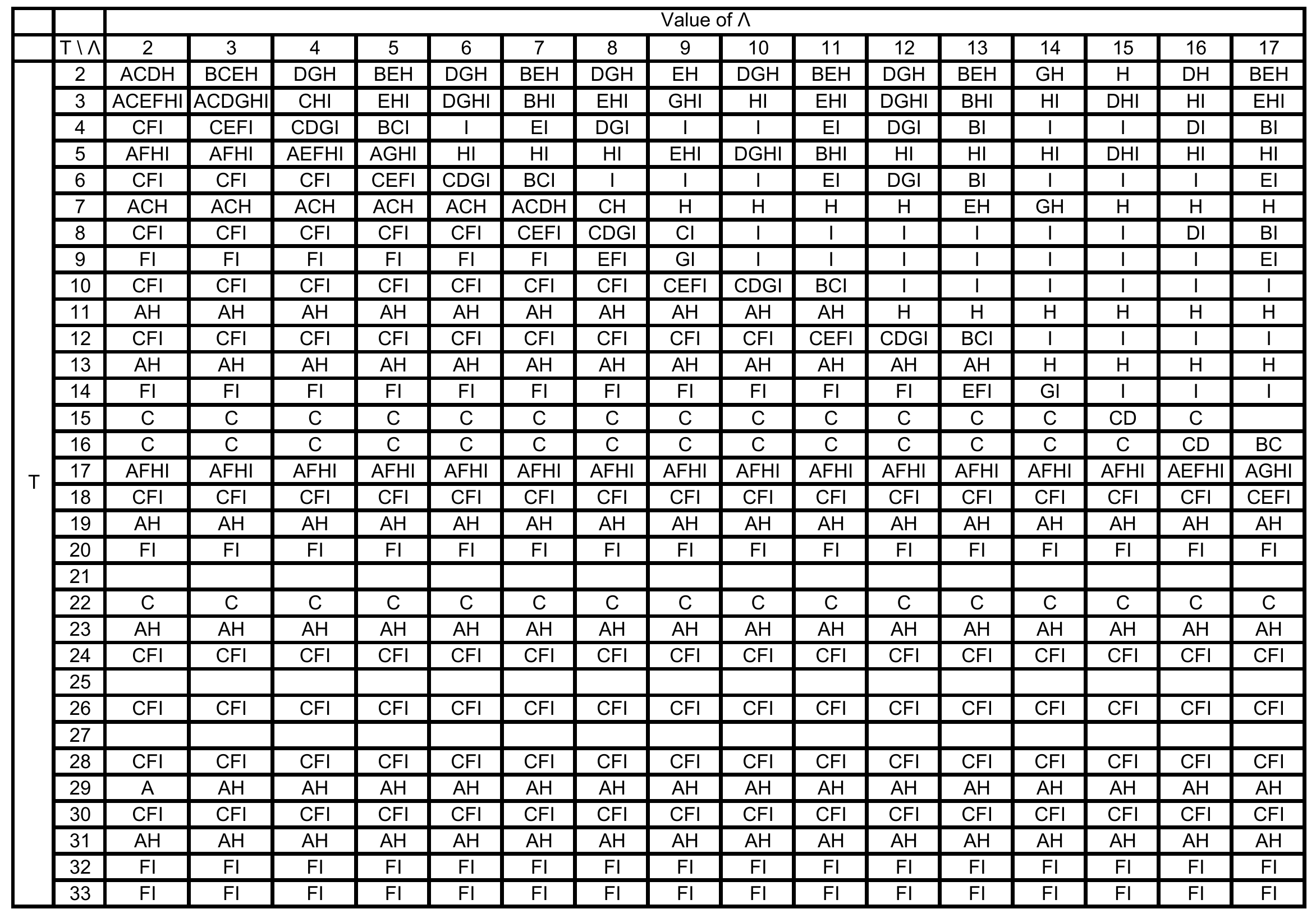}

        \caption{Representative table showing constructions for $ (\Lambda, T), \ \Lambda \in [2,17], \ \ T \in [2,33]$ }
        \label{fig:results}
        \end{center}
\end{figure*}

All the $9$ constructions presented in this paper have been
summarized in Table \ref{2DOOC_table}. Additionally,
Fig.~\ref{fig:results} provides a visual depiction of the parameter
sets for small $\Lambda, T$ in the range, $ 2 \leq \Lambda \leq 17$,
$2 \leq T \leq 33$ for which constructions are now available as a
result of the $9$ constructions presented here.  All constructions
are either optimal or else drawn from a family of asymptotically
optimal constructions and correspond in every case to a code whose
size is large in relation to the number $T$ of time slots. This
table brings out the need for proposing different constructions
since these 9 constructions are applicable to different values of
$\Lambda$ and $T$. For ease of representation, we use the following
legend:

\vspace{0.1in}

\begin{tabular}{|c|c|c|c|}
  \hline
  Construction & Name in &   Construction & Name in \\
  Name& the Table & Name & the Table \\ \hline \hline
  P1 & A & R1 & F \\  \hline
  P2 & B & R2 & G \\   \hline
  P3 & C & CP1 & H \\   \hline
  P4& D & CR1 & I \\      \hline
  P5 & E &  &  \\     \hline

\end{tabular}

\vspace{0.1in}

For example, consider the entry in the table corresponding to $T=5,
\Lambda =3$. The entry reads AFHI. This means that for $T=5$ and
$\Lambda=3$, we are proposing four optimal (or asymptotically
optimal) constructions in this paper, viz., construction P1, R1, CP1
and CR1.

\begin{note}
Note that the rows corresponding to $T=21, 25$ and $27$ are empty.
However, this does imply that there are no constructions for these
value of $T$ - it simply means that there are no constructions for
$2 \leq \Lambda \leq 17$. For example, for $T=21$, we have
construction P2 ( or B) for $\Lambda = 43, 127, 211$ etc. Similarly,
we have construction P2 for $T = 25, \Lambda = 101, 151, 251$; and
for $T=27, \Lambda = 109, 163, 271$ and so on.
\end{note}


The seven constructions (P1 to P5, R1 and R2) are generated by
regarding a codeword in the 2-D code as the graph of a function. For
example, a codeword in a 2-D AM-OPPW code can be regarded as the
graph of a function $t = f(\lambda)$, $0 \leq t \leq T-1$, $0 \leq
\lambda \leq \Lambda-1$ mapping wavelength to time. Analogously, a
codeword in a 2-D AM-OPPTS OOC can be regarded as the graph of a
function $\lambda = f(t)$ , $0 \leq t \leq T-1$, $0 \leq \lambda
\leq \Lambda-1$ mapping time to wavelength.

The first $5$ of these codes (P1 to P5) are constructed using
polynomials over finite fields. Code P1 is optimal while codes P2 to
P5 are asymptotically optimal.

The next two codes (R1 and R2) are constructed using rational
functions over finite fields. Both these codes are defined for even
values of the MCP and are asymptotically optimal.

Next, we compose P1 with a constant weight binary code and generate
the concatenated code CP1. This code is asymptotically optimal and
is AM-OPPW. We do a similar composition of R1 and a constant weight
binary code to generate CR1, which is also asymptotically optimal.

We present a 3-D code construction using polarization as the third
dimension. This code is constructed using the Chinese Remainder
Theorem.

Finally, we use generalized bent functions to construct a family of
efficient asynchronous phase-encoding sequences for optical CDMA. 


\section{New Bounds on the Code Size} \label{sec:new_bounds}


In this section, we begin by proposing a one-dimensional Johnson
bound on constant weight codes over a non-binary alphabet. The bound
will establish the optimality of some of the constructions that we
propose in the sections that follow.

The codes under consideration are over an alphabet ${\cal A}$ of
size $(T+1)$ containing a distinguished element, which we shall call
$0$.   For codes over such an alphabet, we define the Hamming
correlation between two codewords to be the number of symbol
locations in which the two codewords contain the same {\em non-zero}
symbol.

Let $A_T(\Lambda,\omega, \kappa)$ denote the maximum possible size
of a constant-weight code $\cal C$ over the alphabet ${\cal A}$ of
size $T+1$ of length $\Lambda$, Hamming weight $\omega$, and Hamming
correlation $\leq \kappa$.

%

\vspace*{0.2in}

\begin{prop}[Nonbinary Johnson Bound] \label{GenJohnson}

\[A_T(\Lambda,\omega,\kappa)\leq\left\lfloor\frac{T\Lambda}{\omega}\left\lfloor
\frac{T(\Lambda-1)}{\omega-1}\cdots\left\lfloor
\frac{T(\Lambda-\kappa)}{\omega-\kappa}\right\rfloor\right\rfloor\right\rfloor
.
\]

For the proof, we refer the reader to Appendix~\ref{app:johnson}.


\end{prop}

\vspace*{0.2in}

\begin{note} [Recovering the Binary Johnson Bound]
In the binary case, i.e., when $T+1=2$, the above inequality reduces
to the Johnson bound for constant-weight binary codes.
\end{note}

\begin{note} [Singleton Bound]
For the special case when $\Lambda=\omega$, the above bound
reduces to
\[
A_T(\Lambda,\omega,\kappa) \ \leq \ T^{\kappa+1}
\]
and we have, in fact, recovered the Singleton bound of coding
theory.
\end{note}

We now proceed to apply the non-binary Johnson bound to derive
bounds on AM-OPPW 2-D OOCs.

\begin{prop}[Bound on AM-OPPW Code Size] \label{AM-OPPW-Bound}
The maximum possible size $\Phi(\Lambda \times T,\omega,\kappa)$
of an AM-OPPW 2-D OOC with parameters $(\Lambda \times
T,\omega,\kappa)$ satisfies:
\[\Phi(\Lambda\times T,\omega,\kappa)\leq\left\lfloor\frac{\Lambda}{\omega}\left\lfloor
\frac{T(\Lambda-1)}{\omega-1}\cdots\left\lfloor
\frac{T(\Lambda-\kappa)}{\omega-\kappa}\right\rfloor\right\rfloor\right\rfloor
.\]

For the proof, we refer the reader to Appendix~\ref{app:OPPW}.

\end{prop}

\vspace*{0.2in}

\begin{cor}[Bound on OPPW Code Size] \label{OPPW-Bound} The maximum
possible size $\Phi(\Lambda \times T,\omega,\kappa)$ of an OPPW
2-D OOC with parameters $(\Lambda \times T,\omega,\kappa)$
satisfies:
\[\Phi(\Lambda\times T,\omega,\kappa)\ \leq \
T^\kappa. \]
\end{cor}

It follows from the above that the following constructions are
optimal (note that $\kappa = 1$ here) although this was unknown to
the authors (see Table \ref{tab:lit}): \bit \item the second
construction by Yang and Kwong \cite{YanKwo} and \item the
construction by Shivaleela et. al \cite{ShiSivSel}. \eit

The next few sections will each introduce new constructions of
optimal (or asymptotically optimal) 2-D OOCs.

\section{New Constructions Based on Polynomial Functions} \label{sec:poly}

A codeword in a 2-D AM-OPPW OOC can be regarded as the graph of a
function $t = f(\lambda)$, $0 \leq t \leq T-1$, $0 \leq \lambda \leq
\Lambda-1$ mapping wavelength to time.  Analogously, a codeword in a
2-D AM-OPPTS OOC can be regarded as the graph of a function $\lambda
= f(t)$, $0 \leq t \leq T-1$, $0 \leq \lambda \leq \Lambda-1$
mapping time to wavelength.

Without loss of generality, in the constructions below, we will
identify the $T$ time slots with the set $\mathbb{Z}_T$, which we
label as ${\cal T}$. This will allow us to identify cyclic shifts in
the time domain with $\pmod{T}$ additions in $\mathbb{Z}_T$  . We
will identify the $\Lambda$ wavelengths with subsets ${\cal L}$ of
algebraic structures. Thus, $\mid {\cal T} \mid =T$ and $\mid {\cal
L} \mid =\Lambda$.



All of the constructions in this section employ polynomial functions
whose degree is bounded above by the desired value of the MCP
$\kappa$. In the next section, we use rational functions to
construct 2-D OOCs.

While the constructions below have some elements in common, they
also have their differences. The reason for providing the set of
constructions is to provide constructions for as many $(\Lambda, T)$
parameter sets as possible. The impact of the different
constructions given here can be seen in Fig. \ref{fig:results},
where we identify constructions with the entries in the table.

%
%

\subsection{Construction P1: Mapping Wavelength to Time, OPPW, \ $\omega = \Lambda, \
\Lambda \leq T, \ \kappa < \omega, \ T$ prime}

   Here $T=p,$ ${\cal T}=\mathbb{Z}_p$, ${\cal L} \ \subseteq \ \mathbb{Z}_p$
   and we consider polynomials $f(\lambda)$ over $\mathbb{Z}_p$ of degree $\leq \kappa$
   mapping ${\cal
   L} \rightarrow {\cal T}$, where $p$ is a prime. Let $\varphi$ be a mapping from the set of all such functions to the
code matrices, where the $\Lambda \times T$ code matrix $C$
associated to function $f$ is given by $C(\lambda,t)=1$ iff
$f(\lambda)=t$, where $\lambda \in {\cal L}$.

We first show that $\varphi$ is an injective mapping, i.e., the code
matrices $C_1$ and $C_2$, corresponding to polynomials $f_1$ and
$f_2$ respectively, are equal iff $f_1 = f_2$. This is true since
$C_1 (\lambda, t) = C_2 (\lambda, t) \Leftrightarrow$ $f_1(\lambda)
= f_2(\lambda)$. Since $f_1(\lambda) - f_2(\lambda)$ is a non-zero
polynomial of degree $\leq \kappa$, this equation can have a maximum
of $\kappa$ zeroes. Hence, as $\omega > \kappa $ , these two
functions cannot coincide in all the $\omega$ positions.

Since this code is OPPW, the autocorrelation function for each of
these matrices for non-zero cyclic shifts is $0$ by Remark
\ref{note:autocorr}.

We next declare two polynomials $f(\lambda)$ and $f(\lambda) +
\delta$, where $\delta \in {\mathbb Z}_p$, to be equivalent. This
causes the set of all the code matrices to be partitioned into
equivalence classes, each of size $T$. This results in
$\frac{T^{\kappa + 1}}{T} = T^{\kappa}$ code matrices.

Note first that $C_a(\lambda, t + \tau) = 1 = C_b(\lambda, t)$ iff
$f_a(\lambda) - \tau = f_b(\tau)$. For cross-correlation, consider
$C_a(\lambda, t + \tau)$ and $C_b(\lambda, t)$ from different
equivalence classes corresponding to functions $f_a$ and $f_b$
respectively. Next, the polynomial $f_a(\lambda) - \tau -
f_b(\lambda)$ is non-zero since $f_a$ and $f_b$ belong to different
equivalence classes. Since the polynomial is non-zero, it can have a
maximum of $\kappa$ zeroes (the degree of the polynomial); hence,
the two code matrices $C_a$ and $C_b$ can have a maximum of $\kappa$
collisions.

This results in a
  $(\Lambda \times T,\Lambda,\kappa)$ 2-D OOC of size $T^{\kappa}$. This construction can be seen to be optimal by Corollary
\ref{OPPW-Bound}.

%

\begin{note} The elements of this 2D-OOC can also be regarded as corresponding to
 codewords in a Reed-Solomon code under an appropriate equivalence relation between the codewords.
Let $\Lambda,\kappa, T$ be as above. In particular $T = p,$ where
$p$ is a prime. The $[\Lambda,\kappa]$ Reed-Solomon code ${\cal
C}_{RS}$ may be constructed as follows: let $\{ \alpha_1, \alpha_2,
\ldots, \alpha_{\Lambda}\}$ denote a set of $\Lambda$ distinct
elements drawn from $\mathbb{F}_p$. Let ${\cal P}_{\kappa}$ denote
the set of all polynomials over $\mathbb{F}_p$ of degree $\leq
\kappa$. Set \bea {\cal C}_{RS} & = & \left\{ \left(f(\alpha_1),
\ldots, f(\alpha_{\Lambda})\right) \mid f \in {\cal P}_{\kappa}
\right\}. \eea
Next, partition the set of all the codewords into $p^{\kappa}$
equivalence classes by declaring $\underline{c}_1 \sim
\underline{c}_2$ if $\underline{c}_1 - \underline{c}_2 = \eta
\underline{1}$\ , $\ \eta \in \mathbb{F}_p^*$. Finally, we form a
set ${\cal S}$ by picking precisely one element from each
equivalence class and by associating a $(\Lambda \times T)$ matrix
$A(\lambda,t)$ to each vector $\underline{a} \in {\cal S}$ by
setting
\[
A(\lambda,t)=1  \ \ \mbox{  iff  }  a_{\lambda} \ = \ t, \  \ \ 1 \leq
\lambda \leq \Lambda .
\]
\end{note}

\subsection{Construction P2: Mapping Time to Wavelength, OPPTS, \ $ \Lambda = p, \
p \ \text{prime}, \ T \mid p - 1, \ \omega = T, \ \kappa < \omega$}

Here ${\cal L}=\mathbb{Z}_p$. Let $\alpha$ be an element of
$\mathbb{Z}_p$ of multiplicative order $T$ and let $H$ be the
subgroup of order $T$ generated by $\alpha$ in ${\mathbb{Z}}_p$. We
will identify ${\mathbb{Z}}_T$ with $H$ by associating $t$ with
$\alpha^t$. Consider polynomials $f(x)$ over ${\mathbb{Z}}_p$ of
degree $\leq \kappa$ mapping $H \rightarrow {\cal L}$. Let $\varphi$
be a mapping from the set of all such functions to the code
matrices, where the $\Lambda \times T$ code array $C$ is obtained by
setting
    $C(\lambda,t)=1$ iff $f(\alpha^t)=\lambda$, where $t \in {\cal T} $
    and $ \lambda \in \mathbb{Z}_p$.

We first prove that $\varphi$ is injective, i.e., the code matrices
$C_1$ and $C_2$, corresponding to polynomials $f_1$ and $f_2$
respectively, are equal iff $f_1 = f_2$. This is true since $C_1
(\lambda, t) = C_2 (\lambda, t) \Leftrightarrow$ $f_1(\alpha^t) =
f_2(\alpha^t)$. Since $f_1(\alpha^t) - f_2(\alpha^t)$ is a non-zero
polynomial of degree $\leq \kappa$, this equation can have a maximum
of $\kappa$ zeroes. Hence, as $\omega > \kappa$, these two
polynomials cannot coincide in all the $\omega$ positions.

    We next discard all sub-period polynomials, i.e.,
    polynomials $f(x)$ that satisfy $f(\alpha^i x) = f(x)$ for some $i
    \in \mathbb{Z}_T^*$. This ensures good autocorrelation. Since $f(\alpha^i x) -
    f(x)$ is not the zero polynomial, we know that it has a maximum of $\kappa$ zeroes
    (the degree of the polynomial). Hence, the autocorrelation is bounded above by $\kappa$.

    We now define two polynomials $f(x), g(x)$ to be equivalent
    if $f(\alpha^i x) = g(x)$ for some $i \in {\mathbb{Z}}_T$.
We pick a code matrix corresponding to each equivalence class to
form a code of size $\frac{1}{T}
\sum_{d|(\Lambda-1)}\left(\Lambda^{\left\lceil \frac{\kappa
  +1}{d}\right\rceil}-1\right)\mu(d)$, where $\mu(\cdot)$ is the Mobius function
  (see \cite{MorZhaKumZin} for code size computation).

  Now, consider two polynomials $f_a$ and $f_b$ drawn from different
  equivalence classes. Consider the corresponding code matrices
  $C_a$ and $C_b$. We know that $C_a(\lambda, t + \tau) = C_b(\lambda, t)
  \Leftrightarrow  f_a(\alpha^{t + \tau} ) = f_b(\alpha^t)$. Since these two
  polynomials have been drawn from distinct equivalence classes, the difference polynomial
  $f_a(\alpha^{\tau} x ) - f_b(x)$ is non-zero. This implies that
  $f_a(\alpha^{\tau} x ) - f_b(x)$ has $\leq \kappa$ zeroes. This
  proves that the crosscorrelation is bounded above by $\kappa$.

  This construction is shown to be asymptotically optimal in
Appendix~\ref{app:P2}.




\subsection{Construction P3: Mapping Wavelength to Time, AM-OPPW, \ $\omega = \Lambda - \kappa, \ \kappa < \omega,
\ 1 \le \Lambda \le p^m, \ T=p^m-1, \ p$ prime}

 Let $p$ be prime and $\alpha$ a primitive element of
 $\mathbb{F}_{p^m}$.  Let $T = p^m-1$, ${\cal T} = \mathbb{Z}_T$ and
 ${\cal L} \subseteq \mathbb{F}_{p^m}$. Consider polynomials $f(x)$
 over $\mathbb{F}_{p^m}$ of degree $\leq \kappa$ mapping
 ${\cal L} \rightarrow \mathbb{F}_{p^m}^*$. Let $\varphi$ be a
 mapping from the set of all such functions
 to the code matrices, where the $\Lambda \times T$ code array $C$
 is obtained by setting $C(\lambda,t)=1$ iff $f(\lambda)=\alpha^t$,
 where $t \in {\cal T}$ and $\lambda \in {\cal L}$. For those values
    of $\lambda$ such that $f(\lambda)=0$, the entire row is left
    blank.  Clearly, at most $\kappa$ rows in any matrix can be blank.
    To restore the constant weight property, we arbitrarily delete appropriate
    number of $1$'s to keep the weight equal to $\Lambda-\kappa$ for all
the codewords.

We first prove that $\varphi$ is injective, i.e., the code matrices
$C_1$ and $C_2$, corresponding to polynomials $f_1$ and $f_2$
respectively, are equal iff $f_1 = f_2$. This is true since $C_1
(\lambda, t) = C_2 (\lambda, t) \Leftrightarrow$ $f_1(\lambda) =
f_2(\lambda)$. Since $f_1(\lambda) - f_2(\lambda)$ is a non-zero
polynomial of degree $\leq \kappa$, this equation can have a maximum
of $\kappa$ zeroes. Hence, as $\omega > \kappa$, these two
polynomials cannot coincide in all the $\omega$ positions.

Since this code is OPPW, the autocorrelation function for each of
these matrices for non-zero cyclic shifts is $0$ by Remark
\ref{note:autocorr}.

We define two polynomials $f(x), g(x)$ of degree $\leq \kappa$ to be
equivalent
  if $\alpha^{i} f(x) = g(x)$ for some $i \in \mathbb{Z}_T$.
  In our construction, we choose one polynomial from each equivalence class.
  This gives us a code of size
$\frac{(T+1)^{\kappa+1}-1}{T}$.

Now, consider two polynomials $f_a$ and $f_b$ drawn from different
  equivalence classes. Consider the corresponding code matrices
  $C_a$ and $C_b$. We know that $C_a(\lambda, t + \tau) = C_b(\lambda, t)
  \Leftrightarrow  \alpha^{-\tau}f_a(x ) = f_b(x)$. Since these two
  polynomials have been drawn from distinct equivalence classes, the difference polynomial
  $\alpha^{-\tau}f_a(x ) - f_b(x)$ is non-zero. This implies that
  $\alpha^{-\tau} f_a(x ) - f_b(x)$ has $\leq \kappa$ zeroes. This
  proves that the crosscorrelation is bounded above by $\kappa$.

This construction is proved to be asymptotically optimal in
Appendix~\ref{app:P3}.

\subsection{Construction P4: Mapping Time to Wavelength, AM-OPPTS, \
$\Lambda=p^m-1, \ p  \ \text{prime}, \ \omega = T - \kappa, \ \kappa
< \omega, \ T \mid p^m-1 $}

  Let $p$ be prime. Let $\alpha$ be a primitive element of
 $\mathbb{F}_{p^m}$ and let $\beta$ be a non-zero element in
 $\mathbb{F}_{p^m}$ of multiplicative order $T$. Let $H$ be the
 subgroup of $\mathbb{F}_{p^m}$ generated by $\beta$. We will
 identify $\mathbb{Z}_T$ with $H$ by associating $t$ with $\beta^t$.
  Let ${\cal L} = \mathbb{F}_{p^m}^* = \{1, \alpha, \ldots, \alpha^{p^m-2}\}$ and
 ${\cal T} = \mathbb{Z}_T$. Consider polynomials $f(x)$ over
 $\mathbb{F}_{p^m}$ of degree $\leq \kappa$ mapping $H \rightarrow {\cal
 L}$. Let $\varphi$ be a mapping from the set of all such functions
 to the code matrices, where the $\Lambda \times T$ code array $C$
 is obtained by setting    $C(\lambda,t)=1$ iff
 $f(\beta^t)=\lambda$. For those values
    of $\lambda$ such that $f(\beta^t)=0$, the entire column is left
    blank.  Clearly, at most $\kappa$ columns in any matrix can be blank.
    To restore the constant weight property, we arbitrarily delete an appropriate
    number of $1$'s to keep the weight equal to $T-\kappa$ for all
    the codewords.

We first prove that $\varphi$ is injective, i.e., the code matrices
$C_1$ and $C_2$, corresponding to polynomials $f_1$ and $f_2$
respectively, are equal iff $f_1 = f_2$. This is true since $C_1
(\lambda, t) = C_2 (\lambda, t) \Leftrightarrow$ $f_1(\beta^t) =
f_2(\beta^t)$. Since $f_1(\beta^t) - f_2(\beta^t)$ is a non-zero
polynomial of degree $\leq \kappa$, this equation can have a maximum
of $\kappa$ zeroes. Hence, as $\omega > \kappa $ , these two
polynomials cannot coincide in all the $\omega$ positions.

We next discard all sub-period polynomials, i.e.,
    polynomials $f(x)$ that satisfy $f(\beta^i x) = f(x)$ for some $i
    \in \mathbb{Z}_T^*$. This ensures good autocorrelation. Since $f(\beta^i x) -
    f(x)$ is not the zero polynomial, we know that it has a maximum of $\kappa$ zeroes
    (the degree of the polynomial). Hence, the autocorrelation is bounded above by $\kappa$.

    We now define two polynomials $f(x), g(x)$ to be equivalent
    if $f(\beta^i x) = g(x)$ for some $i \in {\mathbb{Z}}_T$.
We pick a code matrix corresponding to each equivalence class to
form a code of size $\frac{1}{T}
\sum_{d|\Lambda}\left((\Lambda+1)^{\left\lceil \frac{\kappa
  +1}{d}\right\rceil}-1\right)\mu(d)$ \ (see \cite{MorZhaKumZin} for code size computation).

  Now, consider two polynomials $f_a$ and $f_b$ drawn from different
  equivalence classes. Consider the corresponding code matrices
  $C_a$ and $C_b$. We know that $C_a(\lambda, t + \tau) = C_b(\lambda, t)
  \Leftrightarrow  f_a(\beta^{t + \tau}  ) = f_b(\beta^t)$. Since these two
  polynomials have been drawn from distinct equivalence classes, the difference polynomial
  $f_a(\beta^{\tau} x ) - f_b(x)$ is non-zero. This implies that
  $f_a(\beta^{\tau} x ) - f_b(x)$ has $\leq \kappa$ zeroes. This
  proves that the crosscorrelation is bounded above by $\kappa$.

Since the number of codewords is similar to the number of codewords
  for construction P2, the proof for asymptotic optimality is along the
  same lines - see Appendix~\ref{app:P2}.

\subsection{Construction P5: Mapping Time to Wavelength, OPPTS, \
$\Lambda=p^m, \ p \ \text{prime}, \ \omega = T, \ \kappa < \omega, \
T \mid p^m - 1$}

  Let $p$ be prime. Let $\alpha$ be a primitive element of
 $\mathbb{F}_{p^m}$ and let $\beta$ be a non-zero element in
 $\mathbb{F}_{p^m}$ of multiplicative order $T$. Let $H$ be the
 subgroup of $\mathbb{F}_{p^m}$ generated by $\beta$. We will
 identify $\mathbb{Z}_T$ with $H$ by associating $t$ with $\beta^t$.
 Let ${\cal L} = \mathbb{F}_{p^m} = \{0, 1, \alpha, \ldots, \alpha^{p^m-2}\}$ and
 ${\cal T} = \mathbb{Z}_T$. Consider polynomials $f(x)$ over
 $\mathbb{F}_{p^m}$ of degree $\leq \kappa$ mapping $H \rightarrow {\cal
 L}$. Let $\varphi$ be a mapping from the set of all such functions
 to the code matrices, where the $\Lambda \times T$ code array $C$
 is obtained by setting    $C(\lambda,t)=1$ iff
 $f(\beta^t)=\lambda$. Thus, the
 construction here is along the lines of the previous construction
  except that the wavelengths are in 1-1 correspondence with
  {\em all} of $\mathbb{F}_{p^m}$.

We first prove that $\varphi$ is injective, i.e., the code matrices
$C_1$ and $C_2$, corresponding to polynomials $f_1$ and $f_2$
respectively, are equal iff $f_1 = f_2$. This is true since $C_1
(\lambda, t) = C_2 (\lambda, t) \Leftrightarrow$ $f_1(\beta^t) =
f_2(\beta^t)$. Since $f_1(\beta^t) - f_2(\beta^t)$ is a non-zero
polynomial of degree $\leq \kappa$, this equation can have a maximum
of $\kappa$ zeroes. Hence, as $\omega > \kappa$ , these two
polynomials cannot coincide in all the $\omega$ positions.

We next discard all sub-period polynomials, i.e.,
    polynomials $f(x)$ that satisfy $f(\beta^i x) = f(x)$ for some $i
    \in \mathbb{Z}_T^*$. This ensures good autocorrelation. Since $f(\beta^i x) -
    f(x)$ is not the zero polynomial, we know that it has a maximum of $\kappa$ zeroes
    (the degree of the polynomial). Hence, the autocorrelation is bounded above by $\kappa$.

    We now define two polynomials $f(x), g(x)$ to be equivalent
    if $f(\beta^i x) = g(x)$ for some $i \in {\mathbb{Z}}_T$.
We pick a code matrix corresponding to each equivalence class to
form a code of size $\frac{1}{T}
\sum_{d|(\Lambda-1)}\left(\Lambda^{\left\lceil \frac{\kappa
  +1}{d}\right\rceil}-1\right)\mu(d)$ \ (see \cite{MorZhaKumZin} for code size computation).

  Now, consider two polynomials $f_a$ and $f_b$ drawn from different
  equivalence classes. Consider the corresponding code matrices
  $C_a$ and $C_b$. We know that $C_a(\lambda, t + \tau) = C_b(\lambda, t)
  \Leftrightarrow  f_a(\beta^{t+\tau} ) = f_b(\beta^t)$. Since these two
  polynomials have been drawn from distinct equivalence classes, the difference polynomial
  $f_a(\beta^{\tau} x ) - f_b(x)$ is non-zero. This implies that
  $f_a(\beta^{\tau} x ) - f_b(x)$ has $\leq \kappa$ zeroes. This
  proves that the crosscorrelation is bounded above by $\kappa$.

The proof for asymptotic optimality of construction P5 is similar
  to the proof for construction P2 - see Appendix~\ref{app:P2}.

\section{New Asymptotically Optimal Constructions Based on Rational Functions}
\label{sec:rational}

We now use rational functions over $\mathbb{F}_{q}$ to construct two
2-D OOCs. Some properties of rational functions that we use are
summarized in the first subsection; the following two subsections
deal with the two constructions.

\subsection{Cyclic Ordering for the Projective Line}
\label{sub:cyclic_ordering}

%
%

Consider all elements of $\mathbb{F}_q^2$ other than the element $[0
\ \ 0]^T$.  We define an equivalence relation amongst these elements
as follows: \bea [a \ \ b]^T & \sim & [c \ \ d]^T,\eea if for some
$\eta \in \mathbb{F}_q^{*}$, we have $[c \ \ d]^T \ = \ [\eta a \ \
\ \eta b]^T$. This partitions $\mathbb{F}_q^2$ into $(q+1)$
equivalence classes, with each class containing $(q-1)$ elements.
The projective line is obtained by taking precisely one element from
each equivalence class. We denote this by $\pgfq$. Thus, there are
$(q+1)$ ``points'' on the projective line. We will use $${a \brack
b}_{\text{eq}} $$ to
denote the equivalence class containing $${a \brack b}.$$

%
%

We next present a cyclic ordering of the elements of $\pgfq$.



%
%

Let $h(x)=x^2+h_1x+h_0$ with $h_1,h_0\in \mathbb{F}_{q}$ be a
primitive polynomial over $\mathbb{F}_{q}$. Let $\alpha$ be a zero
of this polynomial and
\[H=\left[\begin{array}{cc}
0 &-h_0\\ 1 &-h_1 \end{array}\right]  \] be the associated companion matrix.

\begin{theorem} We claim that \begin{equation} \left\{ \left. \left[ H^i {1 \brack 0}
\right]_{\text{eq}} \ \right| \ 0 \leq i \leq q \right\} = \pgfq.
\label{cyclic_ordering}
\end{equation}

\begin{proof}Note from the definition of the companion matrix that
$$ H^i {1 \brack 0} = {a \brack b} $$ is equivalent
to saying that $$\alpha^i \cdot 1 = a + b \alpha. $$ Thus $$H^i {1
\brack 0} \sim H^j {1 \brack 0},$$  with $j>i$ iff $$ \alpha^i =
\theta \alpha^j$$ for some $\theta \in \gfq^{*},$ i.e., iff $$ (q + 1)
\mid (j - i).$$ Hence, the equivalence classes $$ \left. \left[ H^i
{1 \brack 0} \right]_{\text{eq}} \ \right| \ 0 \leq i \leq q
$$ are all distinct, thus proving the theorem.

\end{proof}
\end{theorem}

\begin{example} We present a cyclic ordering of
$\mathbb{P}^1(\mathbb{F}_3)$. The polynomial $f(x)=x^2+x+2$ is a
primitive polynomial over $\mathbb{F}_3$. Thus
\begin{align}
&H=\left[\begin{array}{cc} 0 &-2\\ 1 &-1
\end{array}\right]=
\left[\begin{array}{cc}
 0 &1\\ 1 &2
\end{array}\right]\Rightarrow&\notag\\
&H{1 \brack 0}={0 \brack 1}, \,\, H^2{1\brack 0}={1
\brack 2}, \,\, H^3{1\brack 0}={2 \brack 2}
\Rightarrow&\notag\\
& \mathbb{P}^1(\mathbb{F}_3)=\left\{{1\brack 0}_{\text{eq}},{0\brack 1}_{\text{eq}},{1\brack
2}_{\text{eq}},{2 \brack 2}_{\text{eq}}\right\} \cdot & \notag
\end{align}

It can easily be checked that $$H^4{1\brack 0}={2 \brack 0} \sim {1
\brack 0},$$ which shows that the ordering is cyclic.
\end{example}

\subsection{Construction R1: Mapping Wavelength to Time, OPPW, \
$T=q+1$, for $q=p^m $, \ $p$ prime, $\Lambda \leq q, \ \omega =
\Lambda, \ \kappa < \omega$, $\kappa = 2\kappa' = 2d$}

Set ${\cal T} \ = \ \pgfq$ and ${\cal L} = \{ \alpha_1, \alpha_2,
\ldots, \alpha_{\lambda}, \ldots, \alpha_{\Lambda} \}\ \subseteq
\gfq$.

Let $f, g$ be polynomials over $\gfq$ and let $\phi$ denote the
mapping given by:
$$ \phi(\alpha_i) = {f(\alpha_i) \brack g(\alpha_i)}_{\text{eq}}
, \ \ 1 \leq i \leq \Lambda.$$ We will regard $\phi$ as a rational
function map because when $g(\alpha_i) \neq 0$ for any $i$, we
equivalently have:
$$\phi(\alpha_i) = {\frac{f(\alpha_i)}{g(\alpha_i)} \brack
1}_{\text{eq}}.$$

Let ${\cal F}_d$ denote the class of rational functions $$ {f(x)
\brack g(x)}$$ over $\gfq$ (and hence over ${\cal L}$ by
restriction) where $f$ and $g$ are polynomials over $\gfq$
satisfying: \ben
\item $f$ and $g$ are both non-zero and both of degree $\leq d$,  \item $f$ and $g$ are relatively prime, i.e., $(f,g)=1$,
\item $f$ is monic, and
\item either $f$ or $g$ must be a non-constant function; equivalently, $\frac{f(x)}{g(x)}$ is not the constant function.
\een

The last condition has been included here since it is mathematically convenient to exclude the constant functions at
this stage and bring them back later. Let $\hat{\mu}(\cdot)$, $\ \hat{\mu}(\cdot): \gfq[x] \rightarrow
\mathbb{Z}$ be the function defined by
 \beq \hat{\mu}(b(x))= \begin{cases}
1, & b(x) = 1, \\
(-1)^r, & b(x) \text{ is the product of $r$ monic}, \\
& \text{irreducible polynomials over } \gfq, \\
0, & \text{else}.
\end{cases} \label{eq:mu-hat} \eeq    Then the number $c_d= \mid {\cal F}_d \mid$ of rational functions in ${\cal F}_d$ can be computed from the results in
\cite{MorZhaKumZin} and is given by:
\[
\mid {\cal F}_d\mid \ = \ \sum_{h(x)} \frac{(q^{d-s+1}-1)^2}{(q-1)} \hat{\mu}(h(x)) \ - \ (q-1)
\]
where the sum is over all monic polynomials $h(x) \in \mathbb{F}_q[x]$ of degree $s \leq d$ and where the last term accounts for the disallowed constant functions. It can be shown~\cite{MorZhaKumZin} that
 \beq c(d)= \begin{cases}
q^{2d+1}-q, &d=1,2,3,4,5,6 \\
\geq q^{2d+1}-\frac{q^{2d-6}}{7}, &d \geq 7.
\end{cases}  \label{eq:counting_rational_fns} \eeq

Let $\varphi$ be a mapping from ${\cal F}_d$ to the code matrices,
where the $\Lambda \times T$ code array is obtained by setting
$C(\lambda, t) = 1$ iff $${f(\alpha_{\lambda}) \brack
g(\alpha_{\lambda})} \sim H^t {1 \brack 0}.$$

We first prove that $\varphi$ is injective, i.e., the code matrices
$C_a$ and $C_b$, corresponding to rational functions
${f_a(\alpha_{\lambda}) \brack g_a(\alpha_{\lambda})}$ and
${f_b(\alpha_{\lambda}) \brack g_b(\alpha_{\lambda})}$ respectively,
are equal iff $f_a(x)=f_b(x)$ and $g_a(x)=g_b(x)$.  Clearly,
the code matrices
$C_a$ and $C_b$ are equal iff for every $1 \leq \lambda \leq \Lambda$,
\beq {f_a(\alpha_{\lambda}) \brack
g_a(\alpha_{\lambda})} = \theta_{\lambda} {f_b(\alpha_{\lambda})
\brack g_b(\alpha_{\lambda})} \text{ for some  } \theta_{\lambda} \in
\gfq^* ,
\label{eq:r1_inj} \eeq
i.e., iff
\bean \Leftrightarrow f_a(\alpha_{\lambda}) = \theta_{\lambda}
f_b(\alpha_{\lambda}) \text{ and } g_a(\alpha_{\lambda}) & = &
\theta_{\lambda} g_b(\alpha_{\lambda}) \ \forall \ \lambda \\
\Leftrightarrow \theta_{\lambda}\left[ f_a(\alpha_{\lambda})
g_b(\alpha_{\lambda}) - f_b(\alpha_{\lambda}) g_a(\alpha_{\lambda})
\right] & = & 0 \ \forall \ \lambda \\
\Leftrightarrow f_a(\alpha_{\lambda}) g_b(\alpha_{\lambda}) -
f_b(\alpha_{\lambda}) g_a(\alpha_{\lambda}) & = & 0 \ \forall \
\lambda \\
\Leftrightarrow f_a(x) g_b(x) - f_b(x) g_a(x) & = & 0 \text{ as
polynomials }\\
\Leftrightarrow f_a(x) g_b(x) & = & f_b(x) g_a(x). \eean

Since $f_a(x)$ is coprime to $g_a(x)$, it must be that $f_a(x) |
f_b(x)$. Similarly, $f_b(x) | f_a(x)$. Since both $f_a(x)$ and
$f_b(x)$ are monic, it implies that $f_a(x) = f_b(x)$ which forces
$g_a(x) = g_b(x)$ and hence proves that $\varphi$ is injective.

Since this code is OPPW, the autocorrelation function for each of
these matrices for non-zero cyclic shifts is $0$ by Remark
\ref{note:autocorr}.

Next, we would like to establish that the collection of matrices $\varphi( {\cal F}_d)$ is closed under cyclic shifts along the time axis.  To show this, it is sufficient to show that if $C(\lambda,t)$ is the code matrix associated to element $ {f(x)
\brack g(x)} \ \in \  {\cal F}_d$, then there exists ${f^{'}(x)
\brack g^{'}(x)} \ \in \ {\cal F}_d$ whose associated code matrix is $C(\lambda, t+1 \pmod{T})$.  Such an element ${f^{'}(x)
\brack g^{'}(x)} \ \in \ {\cal F}_d$ is easily found by setting \[
{f^{'}(x)
\brack g^{'}(x)} \ = \ \mu_0 \left[ \begin{array}{cc} 0 & -h_0
\\
1 & -h_1 \end{array} \right] {f(x) \brack g(x) } \ = \mu_0 { -h_0 g(x)
\brack f(x) -h_1 g(x) }
\]
 where $\mu_0$ is chosen to ensure that $(\mu_0)(-h_0)g(x)$ is monic.

Let us define two elements ${f_a(x) \brack g_a(x)}$ and  ${f_b(x) \brack g_b(x)}$ to be equivalent, i.e.,
\[
{f_a(x) \brack g_a(x)} \sim {f_b(x) \brack g_b(x)}
\]
if
the corresponding code matrices $C_a$ and $C_b$ obtained via the mapping $\varphi$, are cyclic shifts of
each other.  Since $\varphi$ is a $1-1$ mapping and since every code matrix $C(\lambda,t)$ in the image of $\varphi$ has $T$ distinct cyclic shifts, it follows that there are $T$ elements within each such equivalence class.  Our 2D-OCDMA code ${\cal C}$ is then the code obtained by selecting one element of ${\cal F}_d$ from each equivalence class and then applying the mapping $\varphi$.   It follows then that this code is of size $\frac{c(d)}{T}+1$, where the $1$ arises from the
inclusion of the constant function and where $c(\cdot)$ is as
defined in \eqref{eq:counting_rational_fns}.

It remains to establish the cross-correlation bound for the code ${\cal C}$.  Consider two rational functions
${f_a(x) \brack g_a(x)},  {f_b(x) \brack g_b(x)}$ and their
corresponding code matrices $C_a$ and $C_b$.  Let ${f_c(x) \brack g_c(x)}$ be the element of ${\cal F}_d$ associated to a cyclic shift of $C_a(\lambda,t)$ by $\tau$.  Let the value of the cross-correlation be $\nu$ and let ${\cal V}$ denote the collection of  $(t,\lambda) \ \in \ \{1,2,\cdots,\Lambda\} \times \{1,2,\cdots,T\}$ such that
\[
C_a(\lambda,t+\tau \pmod{T}) \ = \ C_b(\lambda,t) \ = \ 1.
\]
This implies that for every $\lambda$ such that $(\lambda,t) \in {\cal V}$,
\beq {f_c(\alpha_{\lambda}) \brack
g_c(\alpha_{\lambda})} = \theta_{\lambda} {f_b(\alpha_{\lambda})
\brack g_b(\alpha_{\lambda})} \text{ for some  } \theta_{\lambda} \in
\gfq^* ,
\label{eq:r1_inj1} \eeq
i.e.,
\bean f_c(\alpha_{\lambda}) = \theta_{\lambda}
f_b(\alpha_{\lambda}) \text{ and } g_c(\alpha_{\lambda}) & = &
\theta_{\lambda} g_b(\alpha_{\lambda})  \\
\Leftrightarrow \theta_{\lambda}\left[ f_c(\alpha_{\lambda})
g_b(\alpha_{\lambda}) - f_b(\alpha_{\lambda}) g_c(\alpha_{\lambda})
\right] & = & 0, \\
\Leftrightarrow f_c(\alpha_{\lambda}) g_b(\alpha_{\lambda}) -
f_b(\alpha_{\lambda}) g_c(\alpha_{\lambda}) & = & 0. \eean
But since the product polynomials $f_c(x)g_b(x)$ and $f_b(x)g_c(x)$ both have degree that is bounded above by $2d=\kappa$, it follows that $\mid {\cal V} \mid \leq \kappa$, thereby establishing that all cross-correlations for any cyclic shift are bounded above by $\kappa$, i.e., that the MCP $\leq \kappa$ and we are done.  This construction is proved to be asymptotically optimal in
  Appendix~\ref{app:R1}.

\subsection{Construction R2: Mapping Time to Wavelength, OPPTS, \
$\omega = T, \ T \mid q-1, \ \Lambda = q + 1$, for $q=p^m $, \ $p$
prime,  $\kappa = 2\kappa' = 2d < \omega$}

We associate with each wavelength
  a distinct element of $\pgfq$.  Unlike in the case of Construction R1, the ordering of elements of $\pgfq$
  is immaterial here. We associate $\beta^t$ with
  the $t$-th time slot, $0 \leq t \leq T-1$, where $\beta \in \mathbb{F}_{q}$
  has multiplicative order $T$, $T | (q-1)$.  As before, we associate elements along the time axis with elements of ${\mathbb Z}_T$.

We restrict our rational functions to once again belong to the set
${\cal F}_{d}$ for $d = \frac{\kappa}{2}$, where ${\cal F}_{d}$ is
as defined in the previous subsection. In this case, we
need to necessarily exclude the constant functions since they lead
to constant autocorrelation function equal to $\omega$ for all time shifts and $\omega > \kappa$.  Thus unlike the case of Construction R1, the last condition in the definition of ${\cal F}_{d}$ is needed here.

Let $\varphi$ be a mapping from ${\cal F}_d$ to the code matrices,
where the $\Lambda \times T$ code array associated to $ {f(x)
\brack g(x)}$ is obtained by setting
$C(\lambda, t) = 1$ iff $${f(\beta^t) \brack g(\beta^t)} \sim
\lambda,$$ where $t \in {\mathbb Z}_T$ and $\lambda \in \pgfq$.

Once again, the first objective is to prove that $\varphi$ is injective, i.e., that the code matrices
$C_a$ and $C_b$, corresponding to rational functions ${f_a(x)
\brack g_a(x)}$ and ${f_b(x) \brack g_b(x)}$
respectively, are equal iff $f_a(x)=f_b(x)$ and $g_a(x)=g_b(x)$.  Clearly, $C_a$ and $C_b$ are equal iff
\[
{f_a(\beta^t) \brack g_a(\beta^t)} = \theta_{t} {f_b(\beta^t) \brack
g_b(\beta^t)} \text{ where } \theta_{t} \in \gfq^* \text{ all } 0 \leq t \leq T-1.
\]

This equation is similar to equation (\ref{eq:r1_inj})
obtained in the previous subsection and hence by arguing in similar fashion, we conclude that $\varphi$ is injective.

If $C(\lambda,t)$ is the code matrix corresponding to rational function ${f(x) \brack g(x)}$, and $\mu_f$ is the coefficient of the highest degree term in $f(\beta^{\tau}x)$, then it is clear that $C(\lambda, t + \tau \pmod{T})$ is uniquely associated to $\frac{1}{\mu_f} {f(\beta^{\tau}x) \brack g(\beta^{\tau}x)}$.
Next, note that if the autocorrelation function of $C(\lambda,t)$ is equal to $\omega$ for some nonzero time shift $\tau$, this must mean that the matrices $C(\lambda,t)$ and $C(\lambda, t + \tau \pmod{T})$ are identical, which in turn can only happen by the $1-1$ nature of $\varphi$, if and only if the functions ${f(x) \brack g(x)}$ and $\frac{1}{\mu_f} {f(\beta^{\tau}x) \brack g(\beta^{\tau}x)}$ are identical.
It follows then that to avoid constructing code matrices $C(\lambda,t)$ with
autocorrelation function equal to $\omega$ for some nonzero time shift $\tau$, we need to discard sub-period
functions, i.e., functions that satisfy \beq{f(\beta^{\tau}x) \brack
g(\beta^{\tau}x)} = \theta {f(x) \brack g(x)} , \ \   \theta \in \gfq^*
\label{eq:r2_avoid}\eeq
for any $1 \leq
\tau \leq T-1 $.
Accordingly, our next objective is to calculate the size of the set resulting from discarding the sub-period functions belonging to ${\cal F}_d$.  Let us first calculate the number of polynomial pairs $(f(x),g(x))$
satisfying
 \ben \item $f$ and $g$ are both non-zero and both of degree $\leq d$,  \item $f(x)$ is monic and \item the pair $(f(x),g(x))$ is not sub-periodic, i.e., that ${f(x) \brack g(x)}$ does not satisfy (\ref{eq:r2_avoid}) for any value of time-shift parameter $\tau$ (this condition subsumes the requirement of elements of ${\cal F}_d$, that either $f(x)$ or $g(x)$ be a non-constant polynomial). \een  We will subsequently
modify this count to ensure that $ {f(x) \brack g(x)}$ satisfies the remaining requirement
that will cause ${f(x) \brack g(x)}$ to belong to ${\cal F}_d$, namely that
\ben
\item[4)] $(f,g)=1$ .
\een
Let $$f(x) = \sum_{i=1}^{r} f_i x^{e_i},$$ where without loss of
generality $0 \leq e_1 < e_2 < \ldots < e_r\leq d$ and $f_r=1$. Similarly, let
$$g(x) = \sum_{j=1}^{s} g_j x^{a_j} $$ where $0 \leq a_1 < a_2 < \ldots < a_s\leq d$. It is straightforward to show that \eqref{eq:r2_avoid} holds iff for some $1 \leq \tau \leq (T-1)$:
\bea \beta^{\tau(e_i -
e_1)} & = & 1 \text{ for } 2 \leq i \leq r, \nonumber \\ \beta^{\tau(a_j - a_1)} & = &
1 \text{ for } 2 \leq j \leq s, \ \text{ and }
\\ \beta^{\tau(e_1 - a_1)} & = & 1. \nonumber \eea  It will be found convenient to partition the different ways in which this can happen as follows:  \bit
\item {\bf Case (A)} $e_1=a_1$  and $r=s=1$.
\item {\bf Case (B)} $e_1=a_1$, either $r>1$ or $s>1$ and
\bea
\text{gcd} \left( \{ e_i - e_1 \}_{i=2}^r, \{ a_j - a_1 \}_{j=2}^s, \ T \right) \nonumber \\ = l \text{ for some } l
> 1. \label{eq:r2_gcd_eq_B} \eea
\item {\bf Case (C)} $e_1>a_1$ and \bea
\text{gcd} \left( \{ e_i - e_1 \}_{i=2}^r, \{ a_j - a_1 \}_{j=2}^s,
(e_1 - a_1), \ T \right) \nonumber \\ = l \text{ for some } l
> 1. \label{eq:r2_gcd_eq_1} \eea
\item {\bf Case (D)} $a_1>e_1$ and
\bea
\text{gcd} \left( \{ e_i - e_1 \}_{i=2}^r, \{ a_j - a_1 \}_{j=2}^s,
(a_1 - e_1), \ T \right) \nonumber \\ = l \text{ for some } l
> 1. \label{eq:r2_gcd_eq_1} \eea
\eit
Our interest is in counting the number of rational functions ${f(x) \brack g(x)}$ corresponding to Cases B, C, D where $l=1$ in each case and will carry out the count for the different cases separately.   Clearly the count for Case C and Case D is the same, so it suffices to obtain a count for Cases B and C.  We begin with the count for Case C.

\begin{figure*}
\bea
|C| & = & \frac{1}{T(q-1)} \sum_{h(x)} \left[ \sum_{l|T} \left\{     2 \sum_{e_1=0}^{d-deg(h(x))} \left\{
q^{\lfloor \frac{d-deg(h(x))-e_1}{l}\rfloor  +1} -1\right\} \sum_{c=1}^{\lfloor\frac{e_1}{l}\rfloor} \left\{
q^{\lfloor \frac{d-deg(h(x))-e_1 + cl}{l}\rfloor +1} -1\right\} + \right. \right. \nonumber \\
& & \left. \left. \sum_{e_1=0}^{d-deg(h(x))} \left\{q^{\lfloor
\frac{d-deg(h(x))-e_1}{l}\rfloor  +1} -1\right\}^2 - (q-1)^2 \left(
d-deg(h(x)) + 1 \right)                    \right\}     \mu(l)
\right] \hat{\mu}(h(x)) .\label{eq:r2_overall_count}\eea
\end{figure*}

{\bf Case C Count} For $l | T$, let us define $u_C(l)$ to be the number of integer
sets corresponding to Case C, i.e., the number of integer sets, $\{e_i \mid 0 \leq e_i \leq d, 1 \leq i \leq r\}$, $\{a_j \mid
0 \leq a_j \leq d, 1 \leq j \leq s\}$ where $e_1>a_1$ and where in addition, \bea \text{gcd} \left( \{ e_i -
e_1 \}_{i=2}^r, \{ a_j - a_1 \}_{j=2}^s, |e_1 - a_1|, \ T \right)
\nonumber \\ = l. \ \label{eq:r2_gcd_eq_l}  \eea Our interest lies
in computing $u_C(1)$ and we will do this using Mobius inversion
\cite{MorZhaKumZin}.

Let $y_C(l)$ be the number of
integer sets $\{e_i \mid 0 \leq e_i \leq d, 1 \leq i \leq r\}$,
$\{a_j \mid 0 \leq a_j \leq d, 1 \leq j \leq s\}$ where $e_1>a_1$ and where in addition, \bit
\item $l | T$, \item $l | (e_i - e_1) \ \forall \ 2 \leq i \leq r$,
\item $ l | (a_j - a_1) \ \forall \ 2 \leq j \leq s$ and  \item $ l
| (e_1 - a_1)$ . \eit   It follows that \beq y_C(l) = \sum_{l^{'}: \
l|l^{'}|T} u_C(l') . \label{eq:yu} \eeq Set \bean
\tilde{y}_C(l) & = & y_C(\frac{T}{l}), \\
\tilde{u}_C(l) & = & u_C(\frac{T}{l}).
\eean
Then we can rewrite \eqref{eq:yu} in the form
\bean
\tilde{y}_C(\frac{T}{l}) & = &  \sum_{l^{'}: \ l|l^{'}|T} \tilde{u}_C(\frac{T}{l'}), \\
& = & \sum_{l^{'} : l^{'} \mid T \text{ and } \frac{T}{l'}|\frac{T}{l}} \tilde{u}_C(\frac{T}{l'}),
\eean
i.e., for any divisor $l$ of $T$,
\[
\tilde{y}_C(l) = \sum_{l^{'}|l} \tilde{u}_C(l^{'}).
\]

Our goal is to compute $u_C(1)=\tilde{u}_C(T)$.  Using Mobius inversion allows us to write
\bea u_C(1) \ = \ \tilde{u}_C(T) & = &  \sum_{l | T}
\tilde{y}_C\left(
 \frac{T}{l}  \right) \mu(l), \label{eq:r2_u} \nonumber \\
& = &  \sum_{l | T}
y_C(l) \mu(l), \label{eq:r3_u}
  \eea where
 $\mu(\cdot)$ is the Mobius function.     The value of $y_C(l)$ can be shown to be given by
\bea
y_C(l) \ & = & \ \nonumber \\ & & \hspace{-1in} \sum_{e_1=0}^{d} \left\{
q^{\lfloor \frac{d-e_1}{l}\rfloor  +1} -1\right\} \sum_{c=1}^{\lfloor\frac{e_1}{l}\rfloor} \left\{
q^{\lfloor \frac{d-e_1 + cl}{l}\rfloor +1} -1\right\} .
\label{eq:r4_y_C} \eea
Since we are interested in counting the number of polynomials ${f(x) \brack g(x)}$ where $f(x)$ is monic, we are interested in the quantity $\frac{u_C(1)}{(q-1)}$ which can be computed by substituting for $y_C$ from \eqref{eq:r4_y_C} into \eqref{eq:r3_u} and then dividing by $(q-1)$.

{\bf Case B Count} An analogous argument can be used to show that
the corresponding expressions for $y_B(l)$, $u_B(1)$ in Case B are
given by \bea y_B(l) & = & \sum_{e_1=0}^{d} \left\{q^{\lfloor
\frac{d-e_1}{l}\rfloor  +1} -1\right\}^2 - (q-1)^2 \left(  d + 1
\right),
\nonumber  \\
u_B(1) & = & \sum_{l | T} y_B(l)\mu(l) \label{eq:r4_y_B} \eea
in which the subtracted second term in the expression for $y_B(l)$ ensures that the instances when both $f(x)$ and $g(x)$ are monomials of the same degree are not counted.

{\bf Overall Count} Putting together Cases B, C, D, we see that the analogous expressions for $y(l)$ and $u(1)$ for the desired overall count $u(1)$ are given by
\bea
y(l) \ = \ \nonumber \\ 2 \sum_{e_1=0}^{d} \left\{
q^{\lfloor \frac{d-e_1}{l}\rfloor  +1} -1\right\} \sum_{c=1}^{\lfloor\frac{e_1}{l}\rfloor} \left\{
q^{\lfloor \frac{d-e_1 + cl}{l}\rfloor +1} -1\right\} \nonumber \\
+ \sum_{e_1=0}^{d} \left\{
q^{\lfloor \frac{d-e_1}{l}\rfloor  +1} -1\right\}^2
- (q-1)^2 \left(  d + 1 \right), \label{eq:r4_y_all} \\
u(1) \ = \ \sum_{l | T} y(l)\mu(l). \label{eq:r3_u_all} \eea The
overall count is computed by substituting for $y$ from
\eqref{eq:r4_y_all} into \eqref{eq:r3_u_all} and then dividing by
$(q-1)$.



We next proceed to modify this count to ensure that $f(x)$ and
$g(x)$ are co-prime.

Let us set
\[
N(d, T) \ = \ \frac{u(d,T,1)}{(q-1)}
\]
 where we have written $u(d,T,1)$ in place of $u(1)$ to emphasize that $u$ is a function of both $d,T$ as well.  In the ensuing, it will be found convenient to keep in mind that for any polynomial $h(x)$, the function ${f(x)h(x) \brack g(x)h(x)}$ is sub-periodic iff the function ${f(x) \brack g(x)}$ is sub-periodic.
Let $h(x)$ be a fixed, monic polynomial over $\mathbb{F}_q$ of degree $s$ and let $M(d, T, h(x))$ denote the number of polynomials satisfying

\ben \item $f$ and $g$ are both non-zero and both of degree $\leq d$,  \item $f(x)$ is monic \item $f(x)=h(x) f^{'}(x)$, $g(x)=h(x) g^{'}(x)$, for some $f^{'}(x)$, $g^{'}(x)$, \item $(f^{'}(x),g^{'}(x))$ are not sub-periodic, i.e., that ${f^{'}(x) \brack g^{'}(x)}$ does not satisfy (\ref{eq:r2_avoid}) for any value of time shift parameter $\tau$.
     \een
      It is not hard to show that $M(d, T, h(x))$ is a function of the polynomial $h(x)$ only through its degree $s$ and that moreover,
 \[ M(d, T, h(x)) \ = \   N(d-s, T) .  \]

 Let us also define $M(d, T)$ to be the set of all polynomials $ {f(x) \brack g(x)}$ satisfying
\ben \item $f$ and $g$ are both non-zero and both of degree $\leq
d$,  \item $f(x)$ is monic \item $(f(x),g(x))$ are not sub-periodic,
i.e., that ${f(x) \brack g(x)}$ does not satisfy (\ref{eq:r2_avoid})
for any value of time shift parameter $\tau$ and in addition
\item $(f,g)=1$ .
     \een
Then we can show that $M(d, T)$ is given by
\bean
M(d, T) & = & \sum_{h(x)} M(d,T,h(x))
\hat{\mu}(h(x)), \\
& = & \sum_{h(x)} N(d-\text{deg}(h(x)),T)
\hat{\mu}(h(x)), \eean
 where the function $\hat{\mu}$ is as defined in \eqref{eq:mu-hat} and where the sum is over all monic polynomials $h(x)
\in \gfq[x]$ of degree $\leq d$.
We can now define equivalence classes on the $M(d, T)$ functions that remain as before by defining the functions
$f(x) \brack g(x)$ and $\frac{1}{\mu_f} {f(\beta^{\tau}x) \brack g(\beta^{\tau}x)}$ to be equivalent and choosing precisely one function from each equivalence class.  Since there are no sub-period functions, there are precisely $T$ elements in each equivalence class and thus the total
number of code matrices is then finally given by \bea \frac{M(d, T)}{T} \label{eq:ra_count}\eea and we have completed our count. This overall count is given in \eqref{eq:r2_overall_count}.

It is relatively easy to verify that the autocorrelation function is bounded above by $2d=\kappa$ for all nonzero time shifts and that the crosscorrelation function is uniformly bounded above by $\kappa$ as well.
The proof for asymptotic optimality for construction R2 is given in Appendix \ref{app:R2}.

\section{New Asymptotically Optimal Constructions based on Concatenation} \label{sec:concatenation}

Consider a $(\Lambda \times T,\ \omega, \ \kappa)$ 2-D OOC ${\cal
C}$ under the requirement that there is at most one pulse per
wavelength (AM-OPPW). We present two asymptotically optimal
constructions that are, in a sense, concatenation of a
constant-weight binary code and an OPPW code. We use this method to
construct two new AM-OPPW codes by composing a constant weight
binary code with code P1 (or R1).

Let ${\cal C}_{cw}$ be a constant weight binary $\{0,1\}$ code of
maximum possible size having the following parameters:
length$=\Lambda$, weight$=\omega$, and maximum inner product between
any two codewords $\leq\kappa$. The size of ${\cal C}_{cw}$ is upper
bounded by the one-dimensional Johnson Bound:
\[
\mid {\cal C}_{cw} \mid \ \leq \
\left\lfloor\frac{\Lambda}{\omega}\left\lfloor\frac{\Lambda-1}{\omega-1}\left\lfloor\frac{\Lambda-2}{\omega-2}\cdots
\left\lfloor\frac{\Lambda-\kappa}{\omega-\kappa}\right\rfloor\right\rfloor\right\rfloor\right\rfloor
.
\]

The idea is to construct a 2-D OOC whose code arrays are partitioned
into $\mid {\cal C}_{cw} \mid$ subsets with each subset associated
to a distinct codeword in ${\cal C}_{cw}$. Consider a codeword in
${\cal C}_{cw}$ where the $1$'s in this binary codeword, appear in
the $\omega$ symbol locations
$\lambda_1,\lambda_2,\cdots,\lambda_w$.  We associate with this
codeword, a maximal collection of 2-D code arrays with MCP $\kappa$
which are such that only the wavelengths associated to rows
$\lambda_1,\cdots,\lambda_w$ contain a pulse. No pulse is sent along
any of the other wavelengths.  For any of the $\mid {\cal C}_{cw}
\mid$ choices of $\omega$ wavelengths, let us use a 2-D OOC ${\cal
C}_{oppw}$ with exactly one pulse per wavelength.

It is easy to see that the composition of these two codes forms a
2-D OOC code with parameters $(\Lambda\times T, \omega, \kappa)$ of
size $\mid {\cal C}_{cw} \mid \cdot \mid {\cal C}_{oppw}\mid$.

Since the new code is AM-OPPW, the autocorrelation is $0$ by Remark
\ref{note:autocorr}.

For crosscorrelation, consider two codewords $C_a$ and $C_b$ from
the new code. There are two possibilities: either both $C_a$ and
$C_b$ have originated from the same codeword (of the constant weight
binary code), or from different codewords. If $C_a$ and $C_b$
correspond to the same codeword in the constant weight binary code,
then they have the same selection of rows and hence, their
crosscorrelation is the same as that of the corresponding codewords
from the 2-D OPPW code, which we know is $\leq \kappa$. If, on the
other hand, $C_a$ and $C_b$ have originated from two different
codewords in the constant weight binary code, then the maximum
number of rows that can overlap in these two codewords is bounded
above by $\kappa$, and each row can contribute a maximum of $1$
collisions; hence, the crosscorrelation is bounded above by
$\kappa$.

\begin{note} One might think of using a Steiner system instead of a
constant weight binary code. The size of a $S(\kappa + 1, \omega,
\Lambda)$ Steiner system (which is the same as a $(\kappa +
1)$-$(\Lambda, \omega, 1)$ t-design) is \beq \frac{\left(
\begin{array}{c} \Lambda  \\ \kappa + 1   \end{array} \right)}{\left(
\begin{array}{c} \omega  \\ \kappa + 1   \end{array} \right)} \cdot  \eeq

This reduces to \beq \frac{\Lambda}{\omega}\left(\frac{\Lambda
-1}{\omega-1}\cdots \left(\frac{\Lambda
-\kappa}{\omega-\kappa}\right)\right) \cdot  \eeq

Note that every Steiner system is also a constant weight binary code
and hence satisfies the Johnson Bound for 1-D codes with equality.
However, Steiner systems do not exist for all possible values of
$\kappa, \Lambda, \omega$. Hence, using constant weight binary codes
provides a more general construction.
\end{note}

\subsection{Construction CP1: AM-OPPW, $\kappa<\omega \leq \Lambda, \ T$ is prime}

For the case when $T$ is prime and $\omega \leq T$, a code ${\cal
C}_{oppw}$ of maximum-possible size $T^{\kappa}$ can be constructed
using construction P1. The overall size of the new 2-D OOC in this
case is given by \[ \mid {\cal C}_{cw} \mid \ T^{\kappa} \ \leq \
T^{\kappa} \
\left\lfloor\frac{\Lambda}{\omega}\left\lfloor\frac{\Lambda-1}{\omega-1}\left\lfloor\frac{\Lambda-2}{\omega-2}\cdots
\left\lfloor\frac{\Lambda-\kappa}{\omega-\kappa}\right\rfloor\right\rfloor\right\rfloor\right\rfloor
. \]

\begin{prop} \label{prop:concatenate} In the above construction, the resulting AM-OPPW
2-D OOC is asymptotically optimal, if the constant weight code is
asymptotically optimal (in Johnson Bound).

For the proof, we refer the reader to
Appendix~\ref{app:concatenate}.

\end{prop}


\begin{example}
In this example, we construct a $(7\times 5,3,1)$ AM-OPPW 2-D OOC.
We first need to choose a constant weight code of length $7$, weight
$3$ and $\kappa=1$. We know that the Singer construction
\cite{ChuSalWei,Singer} for OOCs has the desired parameters and that
its corresponding constant weight code is optimal and consists of
the following codewords:
\begin{multline}
\{0,1,3\},\{1,2,4\},\{2,3,5\},\{3,4,6\},\\\{0,4,5\},\{1,5,6\},\{0,2,6\}\notag\end{multline}
Using construction P1 for OPPW 2-D OOC, we construct a $(3\times
5,3,1)$ OPPW 2-D OOC of size $5$. Concatenating these two codes will
result in a $(7\times 5,3,1)$ AM-OPPW 2-D OOC of size 35, which is
optimal by Proposition \ref{AM-OPPW-Bound}.
\end{example}

\subsection{Construction CR1: AM-OPPW, $\omega \leq \Lambda, \ \kappa<\omega$
is even, $T = p^m + 1$}

In a manner similar to the one described above, we compose the OPPW
construction R1 with a constant weight binary code. The overall size
of the new 2-D OOC is given by \[ \mid {\cal C} \mid \leq \left(
\frac{c(\kappa')}{T} + 1 \right)
\left\lfloor\frac{\Lambda}{\omega}\left\lfloor\frac{\Lambda-1}{\omega-1}\left\lfloor\frac{\Lambda-2}{\omega-2}\cdots
\left\lfloor\frac{\Lambda-\kappa}{\omega-\kappa}\right\rfloor\right\rfloor\right\rfloor\right\rfloor
. \]

\begin{prop} \label{prop:concatenate_R1} In the above construction, the resulting AM-OPPW
2-D OOC is asymptotically optimal, if the constant weight code is
asymptotically optimal (in Johnson Bound).

\begin{proof}
Using equation \eqref{eq:R1} of Appendix \ref{app:R1}, we see that
\beq \lim_{\Lambda, T \rightarrow \infty} |C| \geq T^{\kappa},
\label{eq:CR1} \eeq

which is the same as the number of codewords in construction P1. The
rest of the proof follows from Appendix \ref{app:concatenate}.

\end{proof}
\end{prop}

 \begin{figure}[h]

        \centerline{\includegraphics[width=3in]{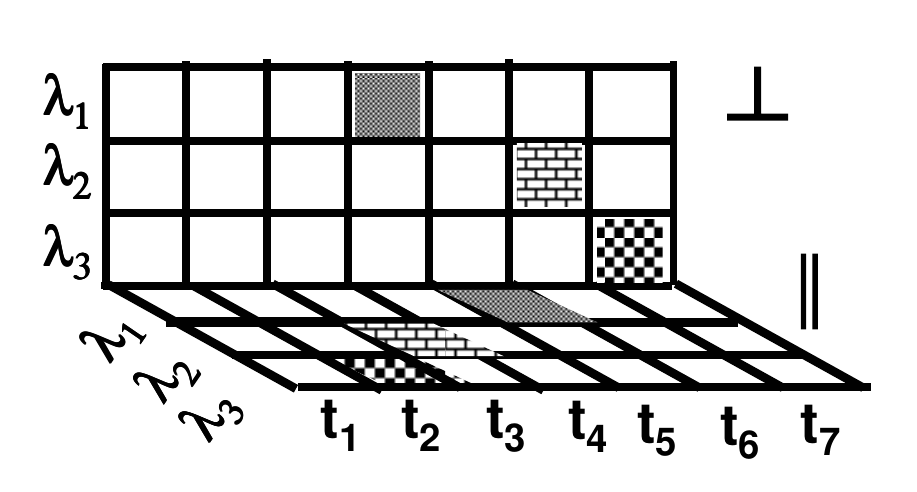}}

\vspace*{0.1in}

        \caption{Example of Polarization Rotation Invariant OOC}\label{fig:3-D}
\end{figure}

\section{Codes Exploiting Polarization} \label{sec:3D}
  A third dimension that can be exploited to construct OOCs is the
  polarization dimension. Light propagates along two orthogonal
  polarization states. Under ideal conditions, these two polarization
  states will be perceived as being orthogonal at the receiver
  \cite{McgMotSagWilOmrKum} despite polarization rotation.

  If we can design codes with good correlation properties that are resilient
   to both timing and polarization ambiguity, we can
  use these codes, which we shall refer to as 3-D OOCs \cite{1362200,kim2000nfs},
  to spread the signal in time, wavelength and
  polarization.

  A 3-D $(2 \times \Lambda\times T,\omega,\kappa)$ OOC  $\cal{C}$ is a family
of $\{0,1\}$ $2 \times \Lambda \times T$ arrays of constant weight
$\omega$. Every pair $\{A,B\}$ of arrays in $\cal{C}$ is required to
satisfy:
        \begin{equation}\label{3-D OOc}
        \sum_{p=1}^{2}\sum_{\lambda=1}^{\Lambda}\sum_{t=0}^{T-1}A(p,\lambda,t)B((p\oplus_2\tau_1),\lambda,(t\oplus_T\tau_2))
        \leq \kappa\\
        \end{equation}
where either $A \neq B$, $\tau_1\neq 0$ or $\tau_2 \neq 0$. All the
bounds of Section \ref{sec:new_bounds} can be generalized to this
class of OOCs.

{\bf Polarization Rotation Invariant Codes Construction}: It is of course
possible to construct a 3-D OOC by starting from a 1-D OOC and
applying the Chinese Remainder Theorem (CRT) \cite{NivZucMon2008}.  When the 3-D code is
constructed from a 2-D code, however, then the transformation can be
seen as a means of reducing the required chip rate.

 Let ${\cal C}$ be a $(\Lambda \times (2T),\omega,
\kappa)$ 2-D OOC where $T$ is an odd integer. For every codeword in
${\cal C}$, apply the CRT mapping to each wavelength to spread that
wavelength in time and polarization. The CRT mapping is a one-to-one
mapping between a sequence and an array that preserves correlation
values.   It follows that by making use of the two orthogonal
polarization states, we have in effect, reduced the needed chip rate
by half.

Fig. \ref{fig:3-D} shows an example of a 3-D OOC with $3$
wavelengths, $7$ time slots, weight $6$ and $\kappa=1$ constructed
using the above construction.

%
%

\section{Sequences for Phase-Encoded OCDMA} \label{sec:phase}


We now focus our attention to designing sequences for phase-encoded
OCDMA.

The spectral encoding OCDMA system is
  harder to implement in comparison with direct-sequence OCDMA.
 That is perhaps the reason why most studies on spectral encoding systems
 have an implementation focus. There is not much literature on
 the subject of spreading sequence design with the exception of a few results on
 spectral amplitude encoding.
 Most experimental results reporting on spectral phase-encoded OCDMA have assumed
 synchronous systems and the
 use of
 Walsh-Hadamard sequences as spreading sequences. As will be shown,
  Walsh-Hadamard Sequences are indeed ideal for the synchronous case but quite unsuitable for
  asynchronous systems.
Other papers in the literature have
 proposed the use of m-sequences or Gold
 sequences as spreading sequences but do not provide adequate justification for their use.

 We first present a model of an asynchronous phase-encoded
 OCDMA system,
 and then identify a metric reflective of the amount of
 cross-correlation (other-user interference) in the system.  Based on this model,
 we formulate the sequence design problem. As will be
 shown, this problem is closely related to the PAPR problem in OFDM
 \cite{DavJed,PatTar,Pat,LitYud}.
 Finally, in the next section, generalized bent functions \cite{KumSchWel} are used to
 construct efficient spreading sequences for an asynchronous system.

\subsection{System Model} \label{sub:system_model}

  \begin{figure}
            \centering
            \includegraphics[width=3.5in]{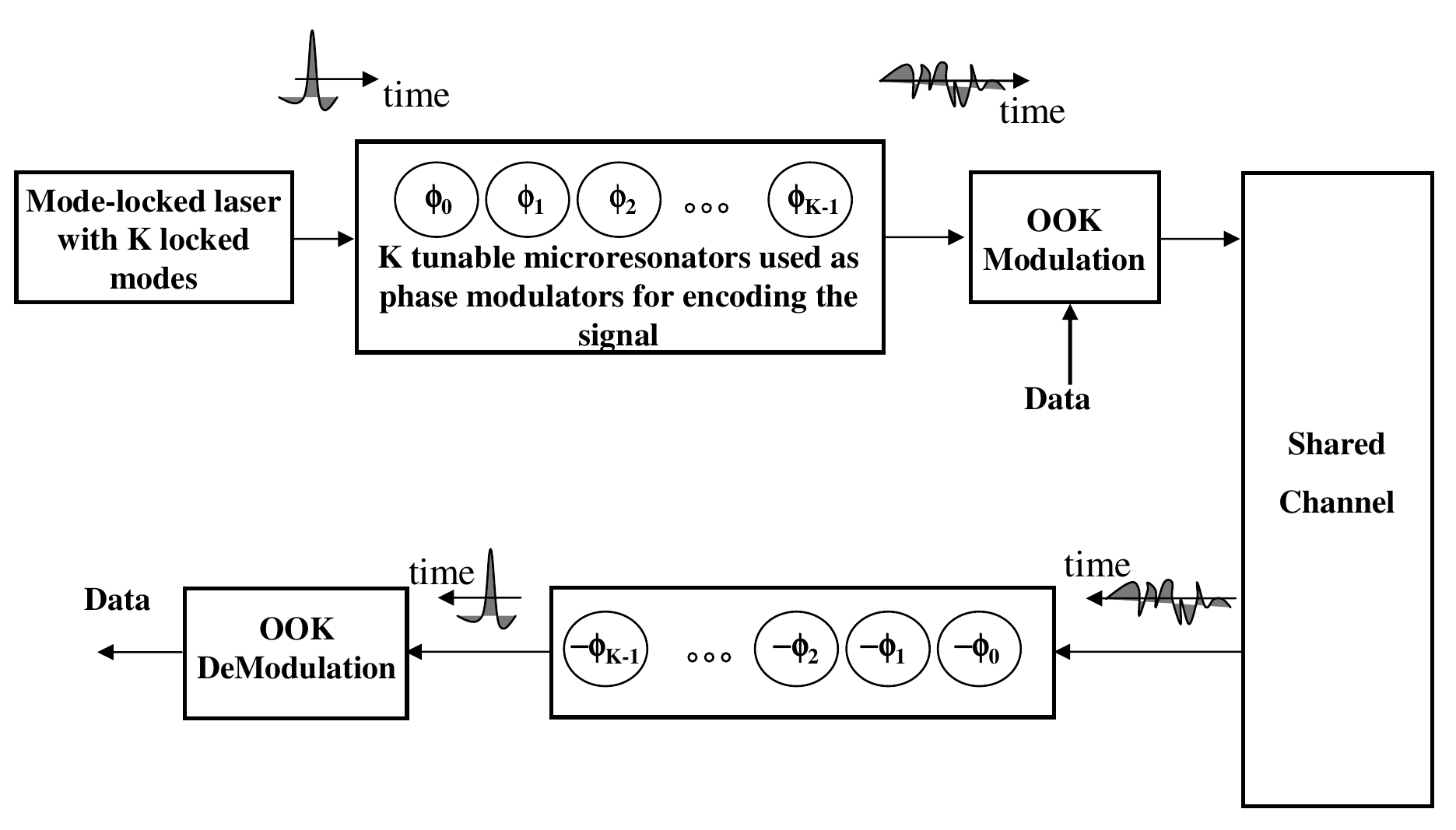}\\
            \caption{Phase Encoding OCDMA system with Coherent Source.}\label{System}
  \end{figure}
  The system that we model in this section is a phase
  encoding OCDMA system with coherent laser source.
  A diagram of this system with a transmitter and receiver is shown in Fig. \ref{System}. The typical laser
  sources used for coherent transmission are mode locked lasers
  (MLL). The electrical field of a mode locked laser can be
  written as
 \bea\label{eq_MLL} E_{MLL}(t)=e^{i\omega_0t}\sum_{k=0}^{K-1}e^{ik(\Delta\omega)t} \;  .\eea
 In this equation, $K$ is the number of modes in the mode locked
 laser, and $\Delta\omega$ is the channel spacing  between two
 consecutive modes in the mode locked laser.

 The output of the MLL is then passed through a phase encoder. In our
 model, the phase encoder applies different phase shifts to
 different modes of the MLL to spread it. Conventionally, the phase masks used in this
 approach consist of only
  $\{0, \pi\}$ phase shifts. Recently, Stapleton et al.
  \cite{StaShaAkhFarPenChoMarObrDap,stapleton2006optical} showed
  that by
  using microdisk resonator technology, any phase can be applied to the
  different modes of the MLL. In light of this result, no restriction on
  the choice of phases is considered in this paper.  The output of the phase encoder is of
  the form
  \bea\label{eq_encoded} E_{Enc}(t)=e^{i\omega_0t}\sum_{k=0}^{K-1}e^{i(k(\Delta\omega)t+\phi_k)} \;  ,\eea
  where $\phi_k$ is the phase shift that the encoder applies to the $k$-th mode of
  the MLL.
Upon OOK modulation with data bit $d$
  \bea\label{eq_transmitted} E_{Tr}(t)=de^{i\omega_0t}\sum_{k=0}^{K-1}e^{i(k(\Delta\omega)t+\phi_k)}, d\in \{0,1\} \;  .\eea
At the receiver,
  the phase decoder applies the inverse phase shift $-\phi_k$ to each mode $k$ of the received signal
 \bea\label{eq_decoded} E_{Dec}(t)&=&de^{i\omega_0t}\sum_{k=0}^{K-1}e^{i(k(\Delta\omega)t+\phi_k-\phi_k)}\notag\\
 &=&de^{i\omega_0t}\sum_{k=0}^{K-1}e^{ik(\Delta\omega)t} \;  ,\eea which is the original
 signal in \eqref{eq_MLL}
modulated by data bit $d$. After the phase decoder,
  a photo detector is used to detect the intensity of the received
 signal
 \bea\label{eq_intensity} \left | E_{Dec}(t) \right | ^2= \left |de^{i\omega_0t}\sum_{k=0}^{K-1}e^{ik(\Delta\omega)t}\right |^2
 =d\left |\sum_{k=0}^{K-1}e^{ik(\Delta\omega)t}\right |^2 \;  .\eea

 If we sample this signal at time $t=0$, then the received signal
 will be $dK^2$ and we can retrieve the transmitted data $d$ using a threshold detector.

 \begin{note}
  In this model, no noise source is considered. This is because we
  wish to focus on the effect of multiple access interference (MAI).
 \end{note}

When there is more than one user transmitting data, the receiver
receives the superposition of the signals.  Assume that users $m$
and $n$ are transmitting data simultaneously and asynchronously.
Each user uses its own phase encoder
 $\Phi^{(\ell)}=\{\phi_0^\ell,\phi_1^\ell,\cdots,\phi_{K-1}^\ell\}$ where $\ell\in
 \{m,n\}$. Let the time difference between user $m$ and $n$ be denoted by $\tau$ ($\tau=0$ in a synchronous system).
The received signal is given by \bean
 E_{Tr}^{(m)}(t)+E_{Tr}^{(n)}(t+\tau) & = &d^{(m)}e^{i\omega_0t}\sum_{k=0}^{K-1}e^{i(k(\Delta\omega)t+\phi_k^{(m)})} + \\
& & \hspace{-1in}
d^{(n)}e^{i\omega_0(t+\tau)}\sum_{k=0}^{K-1}e^{i(k(\Delta\omega)(t+\tau)+\phi_k^{(n)})}
\;  .\eean

The signal at the output of the phase decoder tuned to user $m$
takes on the form
 \bean
& &
\hspace{-.5in}d^{(m)}e^{i\omega_0t}\sum_{k=0}^{K-1}e^{i(k(\Delta\omega)t)}
 \ + \\ & & d^{(n)}e^{i\omega_0
 (t+\tau)}\sum_{k=0}^{K-1}e^{i(k(\Delta\omega)(t+\tau)+(\phi_k^{(n)}-\phi_k^{(m)}))} \;  .
 \eean
The output of  the photo detector of this receiver will be the
square of the magnitude of the above expression. As can be seen,
there is multiple access interference (MAI) at the receiver output.
Each transmitter-receiver pair is assumed to operate synchronously,
and consequently, the receiver samples its output at time $t=0$ to
get \bea \label{eq_output} A=\left |d^{(m)}K
 +d^{(n)}e^{i\omega_0
 \tau}\sum_{k=0}^{K-1}e^{i(k(\Delta\omega)\tau+(\phi_k^{(n)}-\phi_k^{(m)}))}\right
 |^2\notag\\
 =d^{(m)}K^2+d^{(n)}\left |\sum_{k=0}^{K-1}e^{i(k(\Delta\omega)\tau+(\phi_k^{(n)}-\phi_k^{(m)}))}\right
 |^2+\notag\\
 2d^{(m)}d^{(n)}K \re\left(e^{-i\omega_0
 \tau}\sum_{k=0}^{K-1}e^{-i(k(\Delta\omega)\tau+(\phi_k^{(n)}-\phi_k^{(m)}))}\right)
  \;  .\eea
 When $d^{(n)}=0$, there is no interference and we are back to the single-user case.
 Hence, we assume  $d^{(n)}=1$ from now.

\begin{note}
 In the synchronous $\tau=0$ case, if $\Phi^{(m)}$ and $\Phi^{(n)}$
 are Walsh-Hadamard sequences, (i.e., each $\{\exp(i\phi_k^{(\ell)})\}$ is a sequence in a
  Walsh-Hadamard sequence family), the two summations in
\eqref{eq_output} become zero, and there is no
 interference.   This is of course clearly not the case when $\tau\neq 0$
 (see Example \ref{Walsh} and Fig. \ref{Walsh-Hadamard}).
 \end{note}
Setting
\bea\Theta_{nm}(\tau)=\sum_{k=0}^{K-1}e^{-i[k(\Delta\omega)\tau+(\phi_k^{(n)}-\phi_k^{(m)})]}
\;  ,\eea and noting that
 \bea
\mid \re\left(e^{-i\omega_0
 \tau}\Theta_{nm}(\tau)\right) \mid \leq
 \left | \Theta_{nm}(\tau) \right |  \;  ,\eea
we obtain
 \bea
  d^{(m)}K^2+ \left | \Theta_{nm}(\tau) \right |^2-2d^{(m)}K\left | \Theta_{nm}(\tau) \right
  |\leq A \notag\\
  \leq d^{(m)}K^2+ \left | \Theta_{nm}(\tau) \right |^2+2d^{(m)}K\left | \Theta_{nm}(\tau) \right
  |  \;  . \eea
It follows that
minimization of $\left | \Theta_{nm}(\tau) \right|$ for all $\tau$
is a reasonable criterion for signal design.

\begin{note}
The above generalizes in a straightforward fashion to the case of
more than $2$ users.
\end{note}

\subsection{Connection with PAPR problem} \label{sub:papr}

Our objective is thus the design of sequences of length $K$ such
that: \bea\max_{\tau \in [0,\frac{2\pi}{\Delta\omega})}\left
|\sum_{k=0}^{K-1}e^{-i[k(\Delta\omega)\tau+(\phi_k^{(n)}-\phi_k^{(m)})]}\right
| \;  \eea is minimized for every sequence pair $\{\phi_k^{(n)}\},
\{\phi_k^{(m)}\}$.  Equivalently, we seek to minimize \bea\max_{\tau
\in [0,1)}\left
|\sum_{k=0}^{K-1}e^{-ik2\pi\tau}e^{-i(\phi_k^{(n)}-\phi_k^{(m)})]}\right
|  \;  .\eea

The design of sequences with minimum PAPR (peak to
 average power ratio) crops up in conjunction with signal design for OFDM systems
 \cite{litsyn_book,DavJed,PatTar,Pat,LitYud}.  Since designing for low PAPR is
 hard, the common design approach is to
design for low PMEPR (peak-to-mean envelope power ratio), which is
more tractable. The PMEPR problem (see \cite{LitYud}) is one of
designing sequences $\{a_k\}$ that minimize : \bea\max_{\tau \in
[0,1)}\left |\frac{1}{K}\sum_{k=0}^{K-1}a_ke^{-ik2\pi\tau}\right |^2
  \;  .\eea
As can be seen, in our problem, we are interested in phase sequences
$\Phi^{(m)}$ and $\Phi^{(n)}$ such that
$\exp(-i(\Phi^{(n)}-\Phi^{(m)}))$ is a sequence with good PMEPR.

The results in \cite{LitYud} as applied to the present situation are
stated below.  Let  \bea M_d^{(K)}=\max_{j=0,\cdots,K-1}\left
|\sum_{k=0}^{K-1}e^{-ik2\pi\left(\frac{j}{K}\right)}e^{-i(\phi_k^{(n)}-\phi_k^{(m)})]}\right
| \;  \eea and \bea  M_c^{(K)}=\max_{\tau\in [0,1)}\left
|\sum_{k=0}^{K-1}e^{-ik2\pi\tau}e^{-i(\phi_k^{(n)}-\phi_k^{(m)})]}\right
|  \;  .\eea  From \cite{LitYud}, we know
\begin{theorem}\label{prop_1}
$M_d^{(K)}\geq \sqrt{K}\;  .$
\end{theorem}
\begin{theorem}\label{prop_2}
For $K>3$: \bea \hspace{-0.5in} \frac{2}{\pi}\ln
K+0.603-\frac{1}{6K}& < &
\max_{F_K(t)}\left\{\frac{M_c(F_K)}{M_d(F_K)}\right\} \nonumber \\
&  <& \frac{2}{\pi}\ln K+1.132+\frac{3}{K}  \;   \eea

in which

\bea F_K(t) =\sum_{k=0}^{K-1}a_k e^{2\pi ikt} \quad \text{such
that} \quad \sum_{k=0}^{K-1}|a_k|^2 = K \eea

and

\bea M_d(F_K) & = & \max_{j=0,\ldots,K-1}\left| F_K \left(\frac{j}{K}
\right)\right | \ , \nonumber \\
M_c(F_K) & = & \max_{t \in [0,1)}\left| F_K \left(t \right)\right |. \notag \eea

\end{theorem}

The implication of Theorem \ref{prop_2} is that, if we design
sequences with good asynchronous properties for all the
$\frac{j}{K}$ samples of $\tau$, it is guaranteed that the same
sequence has good asynchronous properties for all values of $\tau$.

\section{Sequence Constructions based on Bent Functions} \label{sec:phase_construction}

In this section we use generalized bent functions to design
sequences with good asynchronous properties. Some preliminaries on
generalized bent functions that we will use are introduced in the
first subsection.

  \begin{figure}
            \centering
            \includegraphics[width=3in]{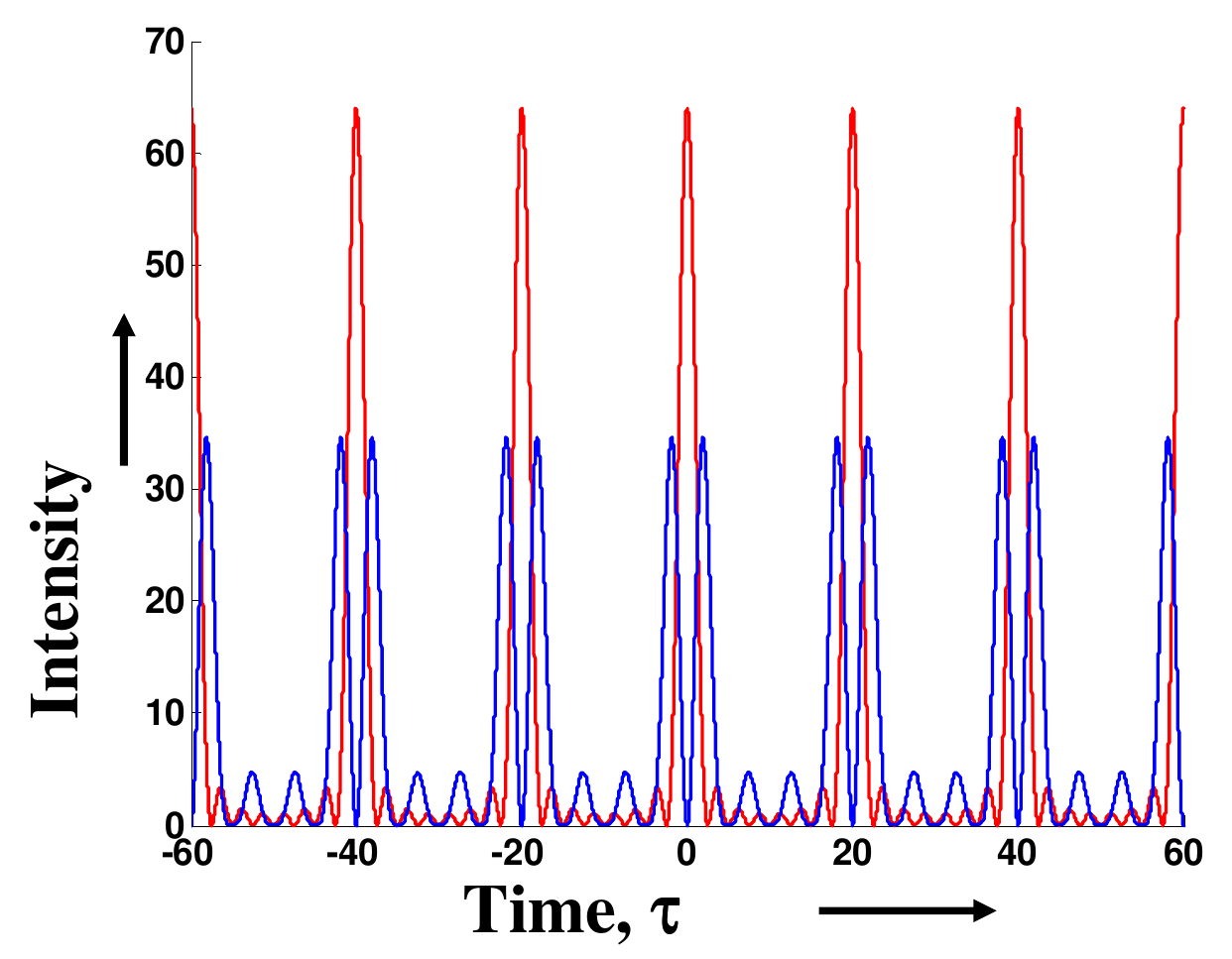}\\
            \caption{An Example of Walsh-Hadamard Sequences with $K=8$.} \label{Walsh-Hadamard}
  \end{figure}

\subsection{Generalized Bent Functions} \label{sub:bent}

\begin{defn}\cite{KumSchWel}
Let $\mathbb{Z}_{q}^m$ denote the set of $m$-tuples with elements
drawn from the set of integers modulo $q$,
$w=e^{i\left(\frac{2\pi}{q}\right)}$ and $g$ a complex-valued
function defined on $\mathbb{Z}_{q}^m$. The Fourier transform of $g$
is then defined to be the function $G$ given by: \bea
G(\lambda)=\frac{1}{\sqrt{q^m}}\sum_{x\in
\mathbb{Z}_{q}^m}g(x)w^{-\lambda^tx}, \quad \lambda \in
\mathbb{Z}_{q}^m \;  .\eea
\end{defn}
\begin{defn}\cite{KumSchWel}
A function $f$, $f:\mathbb{Z}_{q}^m\rightarrow \mathbb{Z}_{q}$ is
said to be bent if all the Fourier transform coefficients of $w^f$
have unit magnitude.
\end{defn}
\begin{theorem}\label{bent_1}\cite{KumSchWel} Every affine or linear translate of a bent function is also bent.
\end{theorem}
\begin{theorem}\label{bent_2}\cite{KumSchWel} Let $q$ be odd. Then the function $f$
over $\mathbb{Z}_{q}$ defined by:
 \bea f(k)=k^2+ck \quad \text{all} \quad k\in \mathbb{Z}_{q} \;  .\eea
 is bent for all $c \in \mathbb{Z}_{q}$.
\end{theorem}
\subsection{New Construction 1} \label{sub:new_1}
\begin{prop}\label{main}
Let \[\Phi=\left\{\Phi^{(\ell)}\mid
\Phi^{(\ell)}=\{\phi_0^\ell,\phi_1^\ell,\cdots,\phi_{K-1}^\ell\},\ell\in\{1,2,\cdots,L\}\right\}\]
be a family of phase sequences such that the difference sequence is
associated to a bent function, i.e.,
\[\Phi^{(n)}-\Phi^{(m)}=\frac{2\pi}{K}(f(0),f(1),\cdots,f(K-1)),
n\neq m\] where $f(x)$ is a bent function over $\mathbb{Z}_{K}$. Then $\max
\left | \Theta_{nm}(\tau) \right|$ is as small as it can possibly be
over multiples $\tau$ of $\frac{2\pi}{K(\Delta\omega)}$, and thus
these phase sequences are suitable for use in asynchronous
phase-encoded OCDMA systems.

For the proof, we refer the reader to Appendix~\ref{app:bent_1}
\end{prop}

\begin{prop} \label{thm:main}
The following phase sequences have the minimum possible value of
$\max \left | \Theta_{nm}(\tau) \right|$ property to be used for
asynchronous phase encoding OCDMA systems with $K$ modes, where $K$
is an odd prime:
\begin{multline} \phi_k^{(m)}=(k^3+a_mk^2+b_mk+c_m)\frac{2\pi}{K},
\\
a_m,b_m,c_m\in \mathbb{Z}_K;\,m\neq n:a_m\neq a_n;\, K \,\text{a
prime}>2.\notag
\end{multline}

For the proof, we refer the reader to Appendix~\ref{app:bent_2}
\end{prop}

  \begin{figure}
            \centering
            \includegraphics[width=3in]{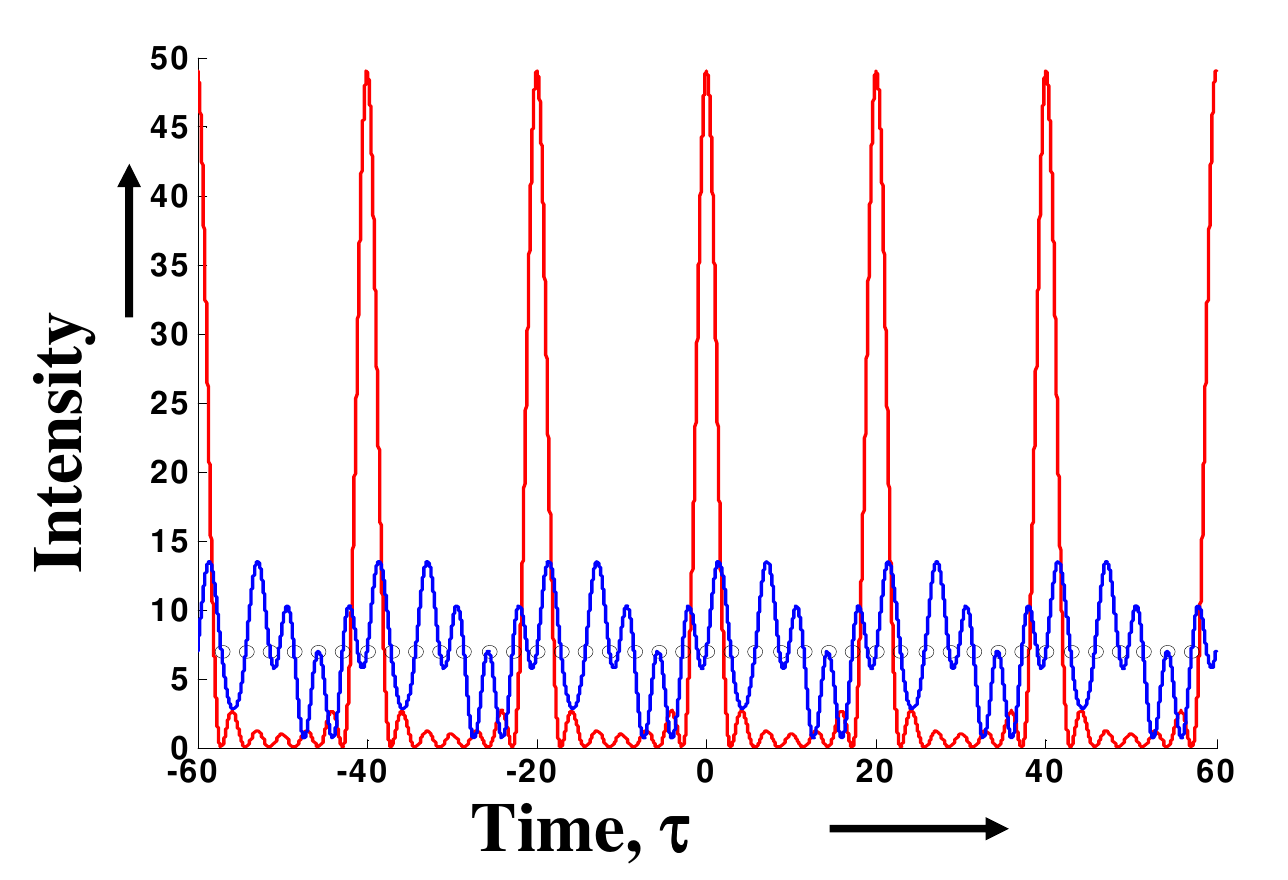}\\
            \caption{An Example of application of Proposition \ref{main} with $K=7$.}\label{bent}
  \end{figure}

  \begin{example}\label{Walsh}
  Fig. \ref{Walsh-Hadamard} shows the application of
  Walsh-Hadamard sequences for asynchronous systems.
In this graph,  $K=8$, $\omega_0=\frac{\pi}{4}$, $\Delta
\omega=\frac{\pi}{10}$, $\Phi^{(m)}=(0,0,0,0,0,0,0,0)$ and
$\Phi^{(n)}=(\pi,\pi,\pi,\pi,0,0,0,0)$. In this figure, the output
of the MLL as it is seen after the photo detector is denoted by the
graph that has a maxima above $60$ (this curve is red colored in the
soft copy). The other curve with a maxima just above $30$ (which is
blue colored in the soft copy) shows $\left | \Theta_{nm}(\tau)
\right|^2$ at the output. As can be seen, the system has no
interference for $\tau=0$ (synchronous case). However, even for
small deviations from $\tau=0$, $\left | \Theta_{nm}(\tau)
\right|^2$ increases significantly. Because of the high peak of
$\left | \Theta_{nm}(\tau) \right|^2$, these phase sequence are not
suitable for asynchronous transmission.
\end{example}

\begin{example}
  Fig. \ref{bent} shows the application of the construction of
Proposition \ref{thm:main}, where $K=7$, $\omega_0=\frac{\pi}{4}$,
$\Delta \omega=\frac{\pi}{10}$, $(a_m,b_m,c_m)=(2,5,3)$ and
$(a_n,b_n,c_n)=(5,4,1)$. For this system
$\Phi^{(m)}=(\frac{6\pi}{7},\frac{8\pi}{7},\frac{2\pi}{7},0,0,0,\frac{12\pi}{7})$
and
\\ $\Phi^{(n)}=(\frac{2\pi}{7},\frac{8\pi}{7},\frac{4\pi}{7},\frac{2\pi}{7},0,\frac{10\pi}{7},\frac{2\pi}{7})$.
In this figure, the output of the MLL as seen after the photo
detector is denoted by the curve that has its maxima around $50$
(red colored in the soft copy). The other curve with maxima around
$15$ (blue colored in the soft copy) shows $\left |
\Theta_{nm}(\tau) \right|^2$ at the output. Here, the circles are
samples of $\left | \Theta_{nm}(\tau) \right|^2$ for
$\tau=\frac{2\pi j}{K(\Delta \omega)}$. As can be seen, all these
values are equal to $7$. It can be observed that $\left |
\Theta_{nm}(\tau) \right|^2$ is low for all values of $\tau$ and the
phase sequences are thus applicable for asynchronous transmission.
 \end{example}

\subsection{New Construction 2} \label{sub:new_2}
The second construction is based on the following family of bent
functions.

\begin{theorem}\label{hab} \cite{42198} Let $q$ be an integer which is neither the product of
distinct primes nor equal to $2$ modulo $4$. Then the function
$f(\cdot)$ over $\mathbb{Z}_q$, defined by
\begin{equation}\label{eq:hab1} f(k+1)=f(k)+ a_k, \ \ \
\forall k, \ a_k \in \mathbb{Z}_q, \ f(0) \in \mathbb{Z}_q\end{equation} is bent if
the integers $a_k$ satisfy the dual conditions:
\begin{equation}\label{eq:hab2}\sum_{k=0}^{K-1} a_k = 0 \pmod{s},\end{equation}
\begin{equation}\label{eq:hab3}a_{k+ns} = a_k + c_1ns \pmod{q}, \ \ \
\forall k \in \mathbb{Z}_s,\forall n \in \mathbb{Z}_{q/s},\end{equation} where $c_1$
is any integer relatively prime to $q$ and $s$ is any integer
greater than one that has the same parity as $q$ and whose square
divides $q$.
\end{theorem}

\vspace{0.2in}

Let $g(k)$ be a second bent function over $\mathbb{Z}_q$ such that
\begin{equation}\label{eq:hab4} g(k+1)=g(k)+ b_k,
\ \ \ \forall k, \ b_k \in \mathbb{Z}_q, \\ g(0) \in \mathbb{Z}_q\end{equation} and
\begin{equation}\label{eq:hab5}\sum_{k=0}^{K-1} b_k = 0 \pmod{s},\end{equation}
\begin{equation}\label{eq:hab6}b_{k+ns} = b_k + c_2ns \pmod{q}, \ \ \
\forall k \in \mathbb{Z}_s,\forall n \in \mathbb{Z}_{q/s},\end{equation} where $c_2$
is any integer relatively prime to $q$.

\begin{prop}\label{prop:hc1} Let $h(k) = f(k) - g(k)$ be a function
over $\mathbb{Z}_q$. Then $h(k)$ is a bent function in
$\mathbb{Z}_q$ if $c_1 - c_2$ is relatively prime to $q$.

For the proof, we refer the reader to Appendix~\ref{app:bent_3}.

\end{prop}

\begin{cor} \label{prop:hc3}Let $q =
\Pi_{j=1}^{n}{p_j}^{r_j}$ such that $p_{min} =
\min{\{p_j\}}_{j=1}^{n}$, where $p_j$ is a prime number. Then the
maximum size is $p_{min}-1$.
\begin{proof} Suppose maximum size $\geq \ p_{min}$. Then $c_i$ and $c_j$ will exist such
that \beqn c_i = c_j \pmod{p_{min}}. \eeqn Hence, we can say that
$p_{min}$ divides $c_i - c_j$. Hence, $c_i - c_j$ is not relatively
prime to $q$, which is a contradiction.\end{proof}
\end{cor}

\begin{cor}\label{prop:hc2}
Let $q = p^r,$ where p is a prime and $r\geq2$. Then the maximum
size is $p-1$.

\begin{proof} Assume that the maximum size $\geq \ p$. Then $c_i$ and $c_j$ will exist such
that \beqn c_i = c_j(mod \ p). \eeqn Hence, we can say that $p$
divides $c_i - c_j$. Hence, $c_i - c_j$ is not relatively prime to
$q$, which is a contradiction.\end{proof}
\end{cor}

  \begin{figure}
            \centering
            \includegraphics[width=3in]{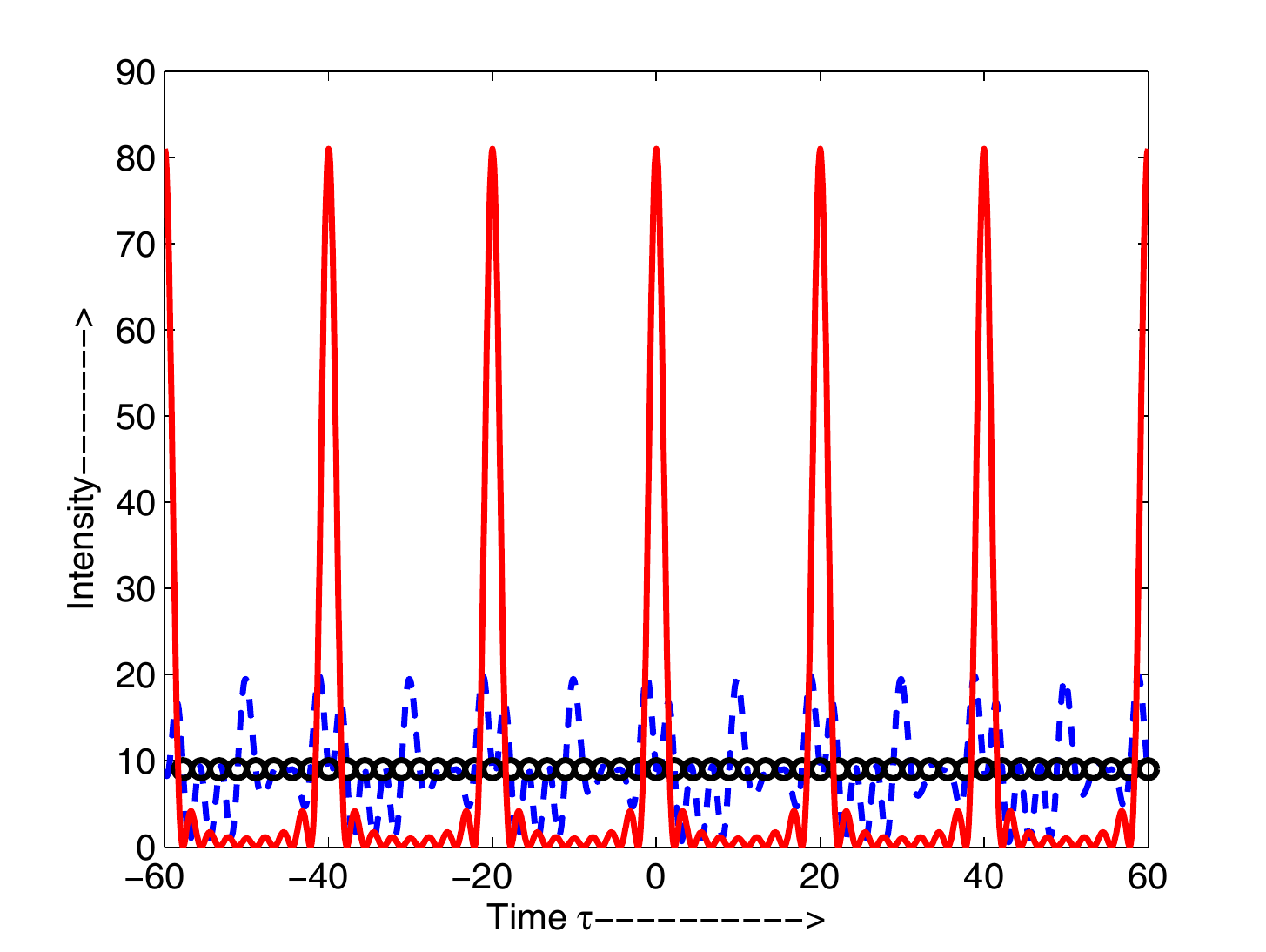}\\
            \caption{An Example of Application of Proposition \ref{prop:hc1} with $q = 9$.}\label{hcp1}
  \end{figure}

\begin{example}
  Fig. \ref{hcp1} shows the application of the construction of
Proposition \ref{prop:hc1}, where $p=3$, $r=2$, $s=2$, $q = p^r =
9$, $\omega_0=\frac{\pi}{4}$, $\Delta \omega=\frac{\pi}{10}$,
$(a_0,a_1,a_2)=(1,3,5)$ and $(b_0,b_1,b_2)=(3,5,7)$. For this
system, $f(0)=1$ and $g(0)=1$. In this figure, the output of the MLL
as it is seen after the photo detector is denoted by the graph that
has its maxima around $80$ (colored red in the soft copy). The
dotted curve with its maxima around $20$ (colored blue in the soft
copy) shows $\left | \Theta_{nm}(\tau) \right|^2$ at the output.
Here, the small circles around $10$ are the samples of $\left |
\Theta_{nm}(\tau) \right|^2$ for $\tau=\frac{2\pi j}{q(\Delta
\omega)}$. As can be seen, all these values are equal to $9$. It can
be observed that $\left | \Theta_{nm}(\tau) \right|^2$ is low for
all values of $\tau$ and hence these phase sequences are good for
asynchronous transmission.
 \end{example}

\section{Conclusion}

In this paper, we presented $9$ families of 2-D OOCs. One of these
families  is optimal and the rest are asymptotically optimal with
respected to the Johnson bound or with respect to the new bounds
proposed in this paper. A novelty of our constructions is the large
size. This was achieved by constructing optimal families for large
values of the MCP since the optimal family size increases
exponentially in the MCP.


\appendices

\section{Nonbinary Johnson Bound} \label{app:johnson}
\begin{center} (Proof of Proposition~\ref{GenJohnson}) \end{center}

The proof is along the lines of the proof of the Johnson bound in
 the case of binary codes.

 Assume a constant weight $(\Lambda,\omega,\kappa)$ code $\mathcal C$ of
 size $A_T(\Lambda,\omega,\kappa)$. If we arrange all the codewords along the rows of a matrix,
 the total weight of the matrix is $\omega A_T(\Lambda,\omega,\kappa)$,
and thus each non-zero symbol is repeated on an average
$\frac{\omega A_T(\Lambda,\omega,\kappa)}{T}$ times. Thus, there is
a symbol $\alpha$ which occurs at least $\frac{\omega
A_T(\Lambda,\omega,\kappa)}{T}$ times, and this symbol is repeated
on an average $\frac{\omega A_T(\Lambda,\omega,\kappa)}{\Lambda T}$
times in each column. It follows that there exists a column $c$ with
at least $\frac{\omega A_T(\Lambda,\omega,\kappa)}{\Lambda T}$
occurrences of the symbol $\alpha$. However, the number of
occurrences of $\alpha$ in column $c$ cannot exceed
$A_T(\Lambda-1,\omega-1,\kappa-1)$. This is because if we select all
the rows containing $\alpha$ in column $c$, and then delete this
column $c$ from all these rows, we will obtain a constant-weight
code of length $\Lambda-1$ and weight $\omega-1$ with Hamming
correlation $\leq \kappa-1$.

It must therefore be that: \bea \label{recursion_1}\frac{\omega
A_T(\Lambda,\omega,\kappa)}{\Lambda T}\leq
A_T(\Lambda-1,\omega-1,\kappa-1)\notag\\ \Rightarrow
A_T(\Lambda,\omega,\kappa)\leq\left\lfloor\frac{\Lambda
T}{\omega}A_T(\Lambda-1,\omega-1,\kappa-1)\right\rfloor . \eea

 By repeating this procedure recursively $\kappa$ times, we arrive at:
\begin{equation}\label{auxineq}
\hspace{-2in} A_T(\Lambda,\omega,\kappa)\leq\end{equation}
{\small{\[\left\lfloor\frac{T\Lambda}{\omega}\left\lfloor\frac{T(\Lambda-1)}{\omega-1}\cdots
\left\lfloor \frac{T(\Lambda-\kappa+1)}{\omega-\kappa+1}
A_T(\Lambda-\kappa,\omega-\kappa,0)\right\rfloor\right\rfloor\right\rfloor
. \]}}

The proof is then completed by noting that
\begin{equation}\label{auxineq1}
 A_T(\Lambda-\kappa,\omega-\kappa,0)\leq
\left\lfloor\frac{(\Lambda-\kappa)T}{\omega-\kappa}\right\rfloor \
.\end{equation}

To see this, let us arrange all the codewords of a
$(\Lambda-\kappa,\omega-\kappa,0)$ constant weight code $\cal{C}$
along the rows of a matrix. To satisfy the constraint of zero
Hamming correlation, each alphabet can occur only once in each
column. Thus, there are at most $T$ non-zero entries in each column.
Since there are $\Lambda-\kappa$ different columns, the entire
matrix can have at most $(\Lambda-\kappa)T$ non-zero entries. On the
other hand, each row has exactly $\omega-\kappa$ non-zero entries,
so that there are at most
$\left\lfloor\frac{(\Lambda-\kappa)T}{\omega-\kappa}\right\rfloor$
rows in the matrix and the assertion is proved.

\section{Bound on AM-OPPW Code Size} \label{app:OPPW}
\begin{center} (Proof of Proposition~\ref{AM-OPPW-Bound}) \end{center}

Let ${\cal C}$ be an AM-OPPW 2-D OOC of size $\Phi(\Lambda\times
T,\omega,\kappa)$. Create a code $\cal {C'}$ that consists of all
$T$ columnar cyclic shifts of each code in $\cal C$. This code is of
size $\Phi'(\Lambda\times T,\omega,\kappa)=T\Phi(\Lambda\times
T,\omega,\kappa)$.

By identifying each row of a code matrix in ${\cal C'}$ having a $1$
in the $t$-th column with the symbol $t$ belonging to $
\{1,2,\ldots,T\}$, and a blank row with the symbol $0$, we obtain
from the 2-D OOC, a 1-D constant weight code over the alphabet
$\mathbb{Z}_{T+1}$.  This 1-D constant weight code has parameters
$(\Lambda,\omega,\kappa)$ and is of size $\Phi'(\Lambda\times
T,\omega,\kappa)$ over an alphabet of size $T+1$.

It follows from our bound above in Proposition \ref{GenJohnson} that
\[\Phi'(\Lambda\times T,\omega,\kappa)\leq \left\lfloor\frac{T\Lambda}{\omega}\left\lfloor
\frac{T(\Lambda-1)}{\omega-1}\cdots\left\lfloor
\frac{T(\Lambda-\kappa)}{\omega-\kappa}\right\rfloor\right\rfloor\right\rfloor\]
\[\Rightarrow \Phi(\Lambda\times T,\omega,\kappa)\leq
\left\lfloor\frac{1}{T}\left\lfloor\frac{T\Lambda}{\omega}\left\lfloor
\frac{T(\Lambda-1)}{\omega-1}\cdots\left\lfloor
\frac{T(\Lambda-\kappa)}{\omega-\kappa}\right\rfloor\right\rfloor\right\rfloor\right\rfloor\]
\[\leq
\left\lfloor\frac{\Lambda}{\omega}\left\lfloor
\frac{T(\Lambda-1)}{\omega-1}\cdots\left\lfloor
\frac{T(\Lambda-\kappa)}{\omega-\kappa}\right\rfloor\right\rfloor\right\rfloor
\ .\]

\section{Asymptotic Optimality of construction P2} \label{app:P2}

The number of codewords in this code are $$\frac{1}{T} \sum_{d |
(\Lambda - 1)} \left( \Lambda^{\left\lceil \frac{\kappa + 1}{d}
\right\rceil} - 1  \right) \mu (d).$$

%
%
%
%
%
%
%

Consider the largest term in the summation - this corresponds to
$d=1$ and evaluates to \beqn \frac{\Lambda^{\kappa + 1} - 1}{T}.
\eeqn

Hence, the total number of codewords $|C|$ is given by
$$|C| = \frac{\Lambda^{\kappa + 1} - 1}{T} + \frac{1}{T} \sum_{d | (\Lambda -
1), \ d > 1} \left( \Lambda^{\left\lceil \frac{\kappa + 1}{d} \right\rceil} -
1  \right) \mu (d).$$

For $\Lambda$ and $T$ tending to infinity, we get

\beq \lim_{\Lambda, T \rightarrow \infty} |C| \ge
\frac{\Lambda^{\kappa+1}}{T}. \label{eq:P2}\eeq

Since $\omega = T$ for this construction, the two-dimensional
Johnson bound given by Theorem~\ref{thm:2-D_bound_A} specializes to
\bean \Phi(\Lambda \times T,\omega,\kappa) & \leq &
\left\lfloor\frac{\Lambda}{T}\left\lfloor\frac{\Lambda
T-1}{T-1}\cdots \left\lfloor\frac{\Lambda
T-\kappa}{T-\kappa}\right\rfloor\right\rfloor\right\rfloor \\
& \leq & \frac{\Lambda}{T}\left(\frac{\Lambda T-1}{T-1}\cdots
\left(\frac{\Lambda T-\kappa}{T-\kappa}\right)\right) . \eean

For $\Lambda$ and $T$ tending to infinity, we get \bea
\lim_{\Lambda,T \rightarrow \infty} \Phi(\Lambda \times
T,\omega,\kappa) & \leq & \frac{\Lambda}{T}\left(\frac{\Lambda
T}{T}\cdots \left(\frac{\Lambda T}{T}\right)\right) \nonumber
\\
& = & \frac{\Lambda^{\kappa+1}}{T}.  \label{eq:JB_P2} \eea

From \eqref{eq:P2} and \eqref{eq:JB_P2}, we can see that
construction P2 is asymptotically optimal. The proof for asymptotic
optimality of constructions P4 and P5 is along similar lines, since
the expression for the total number of codewords is similar.

\section{Asymptotic Optimality of construction P3} \label{app:P3}
The number of codewords $|C|$ in this code are \bean |C| & = &
\frac{(T + 1)^{\kappa + 1}-1}{T} \\
& \ge &  \frac{T^{\kappa + 1}-1}{T}.\eean

When $\Lambda$ and $T$ tend to infinity, we get

\bea\lim_{\Lambda, T \rightarrow \infty} |C| & \ge & \frac{T^{\kappa
+
1}}{T} \nonumber \\
& = & T^{\kappa} .\label{eq:P3} \eea

Since this construction is AM-OPPW, we use the bound given by
Proposition~\ref{AM-OPPW-Bound} to get

\bean \Phi(\Lambda \times T,\omega,\kappa) & \leq &
\left\lfloor\frac{\Lambda}{\omega}\left\lfloor\frac{
T(\Lambda-1)}{\omega-1}\cdots \left\lfloor\frac{
T(\Lambda-\kappa)}{\omega-\kappa}\right\rfloor\right\rfloor\right\rfloor
\nonumber \\
& \leq &
\frac{\Lambda}{\omega}\left(\frac{T(\Lambda-1)}{\omega-1}\cdots
\left(\frac{
T(\Lambda-\kappa)}{\omega-\kappa}\right)\right)\nonumber \\
& = & \frac{\Lambda}{\Lambda -
\kappa}\left(\frac{T(\Lambda-1)}{\Lambda - \kappa -1}\cdots
\left(\frac{T(\Lambda -\kappa)}{\Lambda - \kappa
-\kappa}\right)\right) \eean

since $\omega = \Lambda - \kappa$. When $\Lambda$ and $T$
tend to infinity, we get

\bea \lim_{\Lambda, T \rightarrow \infty} \Phi(\Lambda \times
T,\omega,\kappa) & \leq & \frac{\Lambda}{\Lambda }\left(\frac{T
\Lambda}{\Lambda }\cdots \left(\frac{T \Lambda
}{\Lambda }\right)\right) \nonumber \\
& = & T^{\kappa}. \label{eq:JB_P3}\eea

Comparing \eqref{eq:P3} and \eqref{eq:JB_P3}, we can see that this
construction is asymptotically optimal.

\section{Asymptotic Optimality of construction R1} \label{app:R1}

The number of codewords $|C|$ in this code is

\beqn |C| = \frac{c(\kappa')}{T} + 1, \eeqn where

\beqn c(\kappa')= \begin{cases}
q^{2\kappa'+1}-q, &\kappa'=1,2,3,4,5,6 \\
\geq q^{2\kappa'+1}-\frac{q^{2\kappa'-6}}{7}, &\kappa' \geq 7.
\end{cases}  \eeqn

We consider the two cases separately: $\kappa' \leq 6$ and
$\kappa'\ge7$. For $\kappa' \leq 6$, we have

\bea |C| & = & \frac{c(\kappa')}{T} + 1 \nonumber \\
& \geq & \frac{c(\kappa')}{T} \label{eq:R1_3}  \\
& = & \frac{q^{2\kappa'+1}-q}{T} \nonumber \\
& = & \frac{(T-1)^{2\kappa'+1}-T+1}{T}. \nonumber \eea

In the limit $\Lambda, T$ tending to infinity, we get

\bea \lim_{\Lambda, T \rightarrow \infty} |C| & \geq &  T^{2\kappa'} \nonumber \\
& = & T^{\kappa}. \label{eq:R1_1} \eea

For $\kappa' \geq 7$, we have

\bea |C| & = & \frac{c(\kappa')}{T} + 1 \nonumber \\
& \geq & \frac{c(\kappa')}{T} \label{eq:R1_4}  \\
& \geq & \frac{q^{2\kappa'+1}- \frac{q^{2\kappa' - 6}}{7}}{T} \nonumber \\
& = & \frac{(T-1)^{2\kappa'+1}-\frac{(T-1)^{2\kappa' - 6}}{7}}{T}.
\nonumber \eea

In the limit $\Lambda, T$ tending to infinity, we get

\bea \lim_{\Lambda, T \rightarrow \infty} |C| & \geq &  T^{2\kappa'} \nonumber \\
& = & T^{\kappa}. \label{eq:R1_2} \eea

We can see that \eqref{eq:R1_1} and  \eqref{eq:R1_2} are the same.
Hence, \beq \lim_{\Lambda, T \rightarrow \infty} |C| \geq T^{\kappa}
\label{eq:R1} \eeq for all values of $\kappa'$, i.e., for all even
values of $\kappa$.

The two-dimensional Johnson bound given by
Theorem~\ref{thm:2-D_bound_A} is

\bean \Phi(\Lambda \times T,\omega,\kappa) & \leq &
\left\lfloor\frac{\Lambda}{\omega}\left\lfloor\frac{\Lambda
T-1}{\omega-1}\cdots \left\lfloor\frac{\Lambda
T-\kappa}{\omega-\kappa}\right\rfloor\right\rfloor\right\rfloor
\nonumber \\
& \leq & \frac{\Lambda}{\omega}\left(\frac{\Lambda
T-1}{\omega-1}\cdots \left(\frac{\Lambda
T-\kappa}{\omega-\kappa}\right)\right)\nonumber \\
& = & \frac{\Lambda}{\Lambda }\left(\frac{\Lambda T-1}{\Lambda
-1}\cdots \left(\frac{\Lambda T-\kappa}{\Lambda
-\kappa}\right)\right) \eean

since $\omega = \Lambda$ for this construction. In the limit
$\Lambda, T$ tending to infinity, we get

\beq \lim_{\Lambda, T \rightarrow \infty} \Phi(\Lambda \times
T,\omega,\kappa) \leq T^{\kappa}. \label{eq:JB_R1} \eeq

By comparing \eqref{eq:R1} and \eqref{eq:JB_R1}, we can see that
this construction is optimal.

\section{Asymptotic Optimality of construction R2} \label{app:R2}

The number of codewords $|C|$ in this code is

\begin{eqnarray*} |C| & = & \frac{M(d,T)}{T}  \\
& = & \frac{1}{T} \sum_{h(x)} \hat{\mu}(h(x)) N(d-deg(h(x)),T).
 \end{eqnarray*}

Note that $N(\cdot)$ is an increasing function in its first argument, which gets maximized when the degree of $h(x)$ is $0$. Moreover, for $h(x)=1$, $\hat{\mu}(h(x))$ achieves its maximum value of $1$. Since the summation above is greater than the largest term of the summation (which corresponds to $h(x)=1$), we conclude that

\begin{eqnarray*} |C| & \ge & \frac{1}{T} N(d,T).
 \end{eqnarray*}

Since we are interested in proving the asymptotic optimality, it is sufficient to work with the highest order term. Substituting the value of $N(\cdot)$, the above sum simplifies to

\begin{eqnarray*} |C| & \ge & \frac{u(1)}{T(q-1)}.
 \end{eqnarray*}

From the definition of $u(1)$ in \eqref{eq:r3_u_all}, we notice that its  maximum term corresponds to $l=1$. This is so because $y(\cdot)$ decreases exponentially with $l$. Moreover, $\mu(l)$ attains its maximum value of $1$ for $l=1$. Substituting the value of (the largest term in the summation of) $u(1)$ from \eqref{eq:r3_u_all}  into the above summation, we get

\begin{eqnarray} |C| & \ge & \frac{y(1)}{T(q-1)}. \label{eq:app_R2_1}
 \end{eqnarray}

We now compute an approximation for $y(1)$ and then substitute it back in the above inequality. We do this by substituting $l=1$ in \eqref{eq:r3_u_all} to get the following approximation for $y(1)$

$$ 2 \sum_{e_1=0}^{d} \left\{q^{d-e_1+1} -1\right\} \sum_{c=1}^{e_1} \left\{
q^{d-e_1 + c +1} -1\right\} + \sum_{e_1=0}^{d} \left\{
q^{d-e_1+1} -1\right\}^2.$$

Each summation in this expression is a geometric progression. Summing the series and approximating (by neglecting the lower-order terms in the summation), we get

$$ 2 q^{d+1} q^{d+1} + q^{2d+2}.$$

Substituting this value of $y(1)$ into \eqref{eq:app_R2_1}, we get

\begin{eqnarray} |C| & \ge & \frac{3q^{2d+2}}{T(q-1)} \nonumber\\
   & = &     \frac{3(\Lambda-1)^{\kappa+2}}{T\Lambda} \nonumber \\
   & \ge &   \frac{\Lambda^{\kappa+1}}{T}.   \label{eq:app_R2_2}
 \end{eqnarray}

Comparing this with \eqref{eq:JB_P2}, we see that construction R2 is asymptotically optimal.

\section{Asymptotic Optimality of Construction CP1}
\label{app:concatenate}

\begin{center}  (Proof of Proposition~\ref{prop:concatenate}) \end{center}

\begin{multline}
\left\lfloor\frac{\Lambda}{\omega}\left\lfloor\frac{\Lambda-1}{\omega-1}\left\lfloor\frac{\Lambda-2}{\omega-2}\cdots
\left\lfloor\frac{\Lambda-\kappa}{\omega-\kappa}\right\rfloor\right\rfloor\right\rfloor\right\rfloor\geq\\
\left(\frac{\Lambda}{\omega}\left(\frac{\Lambda-1}{\omega-1}\left(\frac{\Lambda-2}{\omega-2}\cdots
\left(\frac{\Lambda-\kappa}{\omega-\kappa}-1\right)\cdots-1\right)-1\right)-1\right)\\
\Rightarrow \lim_{\Lambda\rightarrow
\infty}\left\lfloor\frac{\Lambda}{\omega}\left\lfloor\frac{\Lambda-1}{\omega-1}\left\lfloor\frac{\Lambda-2}{\omega-2}\cdots
\left\lfloor\frac{\Lambda-\kappa}{\omega-\kappa}\right\rfloor\right\rfloor\right\rfloor\right\rfloor\geq\\
\frac{\Lambda^{\kappa+1}}{\omega(\omega-1)\cdots(\omega-\kappa)}\\
\Rightarrow\lim_{\Lambda\rightarrow \infty}\mid {\cal{C}}_{cw}\mid
T^\kappa \geq
\frac{T^\kappa\Lambda^{\kappa+1}}{\omega(\omega-1)\cdots(\omega-\kappa)}.
\notag
\end{multline}
On the other hand, from the bound in Proposition
\ref{AM-OPPW-Bound}:
\begin{multline}
\Phi(\Lambda\times
T,\omega,\kappa)\leq\left\lfloor\frac{\Lambda}{\omega}\left\lfloor
\frac{T(\Lambda-1)}{\omega-1}\cdots\left\lfloor
\frac{T(\Lambda-\kappa)}{\omega-\kappa}\right\rfloor\right\rfloor\right\rfloor\leq
\\
\frac{\Lambda}{\omega}\frac{T(\Lambda-1)}{\omega-1}\cdots\frac{T(\Lambda-\kappa)}{\omega-\kappa}\leq
\frac{T^\kappa \Lambda(\Lambda-1)\cdots(\Lambda-\kappa)}{
\omega(\omega-1)\cdots(\omega-\kappa)}\\
\Rightarrow \lim_{\Lambda\rightarrow \infty}\Phi(\Lambda\times
T,\omega,\kappa)\leq
\frac{T^\kappa\Lambda^{\kappa+1}}{\omega(\omega-1)\cdots(\omega-\kappa)}\
,\notag
\end{multline}

which establishes what we need to prove.

%
%
%
%

\section{Proof of Proposition \ref{main}} \label{app:bent_1}

\[\left | \Theta_{nm}\left(\frac{2\pi j}{K(\Delta\omega)}\right)\right|=\left|\sum_{k=0}^{K-1}e^{-i[k(\Delta\omega)\frac{2\pi j}{K(\Delta \omega)}+\frac{2\pi f(k)}{K}]}\right|=\]
\[\left|\sum_{k=0}^{K-1}e^{-i\frac{2\pi}{K}jk}e^{-i\frac{2\pi}{K}f(k)}\right|=\left|\sum_{k=0}^{K-1}w^{-jk}w^{-f(k)}\right|=F(j),\]
where $w=e^{i\frac{2\pi}{K}}$. Since the function $f(x)$ is a bent
function:
\[\frac{1}{\sqrt{K}}F(j)=1\Rightarrow
F(j)=\sqrt{K}\]

so for all $j$, $\left | \Theta_{nm}(\frac{2\pi
j}{K(\Delta\omega)})\right|=\sqrt{K}$, and by Theorem \ref{prop_1}
this is the minimum value $\left | \Theta_{nm}(\frac{2\pi
j}{K(\Delta\omega)})\right|$ can take.

By Theorem \ref{prop_2}, this sequence is good for all
values of $\tau$.

\section{Proof of Proposition \ref{thm:main}} \label{app:bent_2}

For any two different sets of phases $\Phi^{(n)}$,$\Phi^{(m)}$ from
the given set of phase sequences, we have
\[\phi_k^{(m)}=
(k^3+a_mk^2+b_mk+c_m)\frac{2\pi}{K}\] and \[\phi_k^{(n)}=
(k^3+a_nk^2+b_nk+c_n)\frac{2\pi}{K}.\] So
\begin{multline}
\phi_k^{(n)}-\phi_k^{(m)}=\notag\\\left((a_n-a_m)k^2+(b_n-b_m)k+(c_n-c_m)\right)\frac{2\pi}{K}.\notag\end{multline}
Since we assume $a_n\neq a_m$, it implies that
\[(a_n-a_m)k^2+(b_n-b_m)k+(c_n-c_m)\] is a bent function by Theorems
\ref{bent_1} and \ref{bent_2}. So the given set of sequences
satisfies the conditions of Proposition \ref{main} and is good to be
used with asynchronous phase encoding sequences.

For any two different set of phases $\Phi^{(i)}$,$\Phi^{(j)}$  from
the above set of phases, we have: $\phi_k^{(i)}=
(k^3+a_ik^2+b_ik+c_i)\frac{2\pi}{K}$ and $\phi_k^{(j)}=
(k^3+a_jk^2+b_jk+c_j)\frac{2\pi}{K}$. Then \bea
\phi_k^{(i)}-\phi_k^{(j)}=\left((a_i-a_j)k^2+(b_i-b_j)k+(c_i-c_j)\right)\frac{2\pi}{K}.\eea
Since we assume $a_i\neq a_j$, it implies that
$(a_i-a_j)k^2+(b_i-b_j)k+(c_i-c_j)$ is a bent function by Theorems
\ref{bent_1} and \ref{bent_2}. Hence, \bea\left
|\frac{1}{\sqrt{K}}\sum_{k=0}^{K-1}
w^{(a_i-a_j)k^2+(b_i-b_j)k+(c_i-c_j)}w^{\lambda k}\right
|=1,\notag\\ w=e^{i\left(\frac{2\pi}{K}\right)}, \lambda
=0,1,\ldots,K-1\notag\\
\Rightarrow\left|\sum_{k=0}^{K-1}e^{i\left[(a_i-a_j)k^2+(b_i-b_j)k+(c_i-c_j)\right]\left(\frac{2\pi}{K}\right)}
e^{i\left(\frac{\lambda2\pi}{K}\right)k}\right|=\sqrt{K}\notag\\
\Rightarrow\left
|\sum_{k=0}^{K-1}e^{i\left(\frac{\lambda2\pi}{K}\right)k}
e^{i(\phi_k^{(i)}-\phi_k^{(j)})}\right|=\sqrt{K}.\eea

So  $\left | \Theta_{nm}(\tau) \right|$ for all $\tau =
\frac{\lambda2\pi}{K}$ is equal to $\sqrt{K}$. Using Theorem
\ref{prop_1} we have met $M_d^{(K)}$. In addition, using Theorem
\ref{prop_2}, we can conclude that $\left | \Theta_{nm}(\tau)
\right|$ is almost equal to $\sqrt{K}$ for all values of $\tau$. So
the sequences described above are good asynchronous phase encoding
OCDMA sequences.

\section{Proof of Proposition \ref{prop:hc1}} \label{app:bent_3}

We know that \bean h(k) & = & f(k) - g(k) \\ \Rightarrow h(k+1) & =
& f(k+1)-g(k+1). \eean

Using equations \eqref{eq:hab1} and \eqref{eq:hab4}, we get

\bean h(k+1) & = & \left( f(k)+a_k \right) - \left( g(k) + b_k
\right) \\ & = & \left( f(k) - g(k)\right) + \left( a_k - b_k
\right). \eean

Hence, we can say that \bean \label{eq:hab7} h(k+1) & = & h(k) +
p_k, \eean where $p_k = a_k - b_k. $

For $h(k)$ to be bent, $p_k$ should satisfy the dual conditions.
Consider

\beqn \sum_{k=0}^{K-1} p_k = \sum_{k=0}^{K-1} (a_k - b_k) \eeqn
\beqn = \sum_{k=0}^{K-1} a_k - \sum_{k=0}^{K-1}b_k. \eeqn

From equations \eqref{eq:hab2} and \eqref{eq:hab5}, we know that
both these sums are equal to $0$, and so is the difference.

We next need to show that $p_k$ satisfies the second condition:
\begin{equation}\label{eq:hab9}p_{k+ns} = a_{k+ns}-b_{k+ns}.\end{equation}

Using equations \eqref{eq:hab3} and \eqref{eq:hab6}, the above
equation can be rewritten as \beqn p_{k+ns} = \left( a_k + c_1ns(mod
\ q)\right) - \left( b_k + c_2ns(mod \ q)\right) \eeqn \beqn =
\left(a_k - b_k\right) + \left(c_1 - c_2\right)ns(mod \ q) \eeqn
\beqn = p_k + \left(c_1 - c_2\right)ns(mod \ q). \eeqn

Hence, $p_k$ satisfies the dual conditions provided that $c_1 \- \
c_2$ is relatively prime to $q$. Hence, $h(k)$ is a bent function.
Hence, the given set of sequences satisfies the condition of
Proposition \ref{main} and is good to be used with asynchronous
phase-encoded OCDMA.

\bibliographystyle{ieeetr}
\bibliography{chapter}

\end{document}